\documentclass[12pt,a4paper]{article}
\usepackage{amsmath,amssymb,epsfig}
\usepackage{epsfig}
\renewcommand{\theequation}{\thesection.\arabic{equation}}
\renewcommand{\thefootnote}{\fnsymbol{footnote}}
\newlength{\extraspace}
\newlength{\extraspaces}
\newcommand{\be}{\begin{equation}
\addtolength{\abovedisplayskip}{\extraspaces}
\addtolength{\belowdisplayskip}{\extraspaces}
\addtolength{\abovedisplayshortskip}{\extraspace}
\addtolength{\belowdisplayshortskip}{\extraspace}}
\newcommand{\ee}{\end{equation}}
 
\newcommand{\ba}{\begin{eqnarray}
\addtolength{\abovedisplayskip}{\extraspaces}
\addtolength{\belowdisplayskip}{\extraspaces}
\addtolength{\abovedisplayshortskip}{\extraspace}
\addtolength{\belowdisplayshortskip}{\extraspace}}
\newcommand{\ea}{\end{eqnarray}}

\newcommand{\bas}{\begin{eqnarray*}
\addtolength{\abovedisplayskip}{\extraspaces}
\addtolength{\belowdisplayskip}{\extraspaces}
\addtolength{\abovedisplayshortskip}{\extraspace}
\addtolength{\belowdisplayshortskip}{\extraspace}}
\newcommand{\eas}{\end{eqnarray*}}
 
\newcounter{subequation}[equation]
\makeatletter

\expandafter\let\expandafter
\reset@font\csname reset@font\endcsname

\def\subeqnarray{\arraycolsep1pt
    \def\@eqnnum\stepcounter##1{\stepcounter{subequation}%
        {\reset@font\rm(\theequation\alph{subequation})}}
\jot5mm     \eqnarray}

\def\subarray{\arraycolsep1pt
    \def\@eqnnum\stepcounter##1{\stepcounter{subequation}%
        {\reset@font\rm(\alph{subequation})}}
\jot5mm     \eqnarray}

\makeatother
\newcommand{\newappendix}[1]{
\vspace{15mm}
\pagebreak[3]
\addtocounter{section}{1}
\setcounter{equation}{0}
\setcounter{subsection}{0}
\renewcommand{\theequation}{\Alph{section}.\arabic{equation}}
\begin{flushleft}
{\large\bf Appendix \Alph{section}: #1}
\end{flushleft}
\nopagebreak
\medskip
\nopagebreak}

\newcommand{\newsection}[1]{
\vspace{15mm}
\pagebreak[3]
\addtocounter{section}{1}
\setcounter{equation}{0}
\setcounter{subsection}{0}
\begin{flushleft}
{\large\bf \thesection. #1}
\end{flushleft}
\nopagebreak
\medskip
\nopagebreak}
 
\newcommand{\newsubsection}[1]{
\vspace{1cm}
\pagebreak[3]
 
\addtocounter{subsection}{1}
\noindent{ \bf \thesection.\arabic{subsection} #1}
\nopagebreak
\vspace{2mm}
\nopagebreak}

\newcommand{\NP}[1]{Nucl.\ Phys.\ {\bf #1}}
\newcommand{\PL}[1]{Phys.\ Lett.\ {\bf #1}}

\newcommand{\PR}[1]{Phys.\ Rev.\ {\bf #1}}

\newcommand{\R}{\mathbb{R}}

\newcommand{\1}{\mbox{1\hspace{-.8ex}1}}
\newcommand{\bra}{\langle}
\newcommand{\ket}{\rangle}
\newcommand{\ra}{\rightarrow}

\newcommand{\rra}{\ \longrightarrow \ }

\newcommand{\is}{ &\! =\! & }
\newcommand{\nonum}{\nonumber \\[1.5mm]}
\newcommand{\sspace}{\makebox[1cm]{ }}

\newcommand{\nspace}{\!\!\!\!\!\!\!\!\!\!}

\newcommand{\inv}{^{-1}}

\newcommand{\lb}{\lambda}
\newcommand{\eps}{\epsilon}

\newcommand{\dd}{{\partial}}

\newcommand{\kbar}{{\overline{k}}}

\newcommand{\Jbar}{{\overline{J}}}

\newcommand{\Gbar}{{\overline{G}}}
\newcommand{\Hbar}{{\overline{H}}}

\newcommand{\cH}{{\cal H}}
\newcommand{\cL}{{\cal L}}

\newcommand{\cV}{{\cal V}}

\newcommand{\mf}{\mathfrak{m}}
\newcommand{\mfb}{\overline{\mathfrak{m}}}
\newcommand{\mft}{\widetilde{\mathfrak{m}}}
\newcommand{\gf}{\mathfrak{g}}
\newcommand{\Gf}{\mathfrak{G}}
\newcommand{\Hf}{\mathfrak{H}}
\newcommand{\Sf}{\mathfrak{S}}

\newcommand{\Bt}{\widetilde{B}}
\newcommand{\Gt}{\widetilde{G}}
\newcommand{\Ht}{\widetilde{H}}
\newcommand{\betat}{\widetilde{\beta}}
\newcommand{\zetat}{\widetilde{\zeta}}
\newcommand{\nt}{\tilde{n}} 
\newcommand{\betabf}{\mbox{\boldmath{$\beta$}}}

\newcommand{\lbbar}{{\overline{\lambda}}}
\newcommand{\hbbar}{{\overline{h}}}

\newcommand{\kappabar}{\overline{\kappa}}
\newcommand{\rhobar}{\overline{\rho}}

\newcommand{\vp}{{\varphi}}
\newcommand{\vpbar}{\overline{\varphi}}

\newcommand{\nbar}{{\bar{n}}}
\newcommand{\dhbar}{\overline{\delta h}}
\newcommand{\sbar}{{\overline{\mathsf s}}}
\newcommand{\ssbar}{\overline{\mathsf S}}

\newcommand{\phihat}{\widehat{\phi}}
\newcommand{\gsf}{\mathsf{g}}
\newcommand{\gsfbar}{\bar{\mathsf{g}}}
\newcommand{\hsf}{\mathsf{h}}
\newcommand{\csf}{\mathsf{c}}
\newcommand{\msf}{\mathsf{m}}
\renewcommand{\b}{\mbox{\tiny \sl B}}

\newcommand{\nl}{\mbox{[\hspace{-.35ex}[}}
\newcommand{\nr}{\mbox{]\hspace{-.32ex}]}}

\begin{document}

%%%%%%%%%%%%%%%%%%%%%%%%%%%%%%%%%%%%%%%%%%%%%%%%%%%%%%%%%%%%%%%%%%%%%%%
\begin{titlepage}
%
%footnotesymbols others than numbers
\renewcommand{\thefootnote}{\fnsymbol{footnote}}
\mbox{}
\vspace{1.5cm}

\begin{center}
{\Large \bf Dimensionally reduced gravity theories}\\[3mm]
{\Large \bf are asymptotically safe}
\vspace{2.5cm}

{{\sc Max Niedermaier}%
\footnote{Membre du CNRS; e-mail: {\tt max@phys.univ-tours.fr}}}
\\[4mm]
{\small\sl Laboratoire de Mathematiques et Physique Theorique}\\
{\small\sl CNRS/UMR 6083, Universit\'{e} de Tours}\\
{\small\sl Parc de Grandmont, 37200 Tours, France}\\
\vspace{2.3cm}

{\bf Abstract}
\vspace{-3mm}

\end{center}

\begin{quote}
4D Einstein gravity coupled to scalars and abelian gauge fields 
in its 2-Killing vector reduction is shown to be quasi-renormalizable 
to all loop orders at the expense of introducing infinitely many 
essential couplings. 
The latter can be combined into one or two functions of the `area 
radius' associated with the two Killing vectors. The renormalization 
flow of these couplings 
is governed by beta functionals expressible in closed form 
in terms of the (one coupling) beta function of a symmetric space 
sigma-model. Generically the matter coupled systems are asymptotically 
safe, that is the flow possesses a non-trivial UV stable fixed point 
at which the trace anomaly vanishes. The main exception is a minimal 
coupling of 4D Einstein gravity to massless free scalars, 
in which case the scalars decouple from gravity at the fixed point. 
\end{quote}
\vfill

\setcounter{footnote}{0}
\end{titlepage}

%%%%%%%%%%%%%%%%%%%%%%%%%%%%%%%%%%%%%%%%%%%%%%%%%%%%%%%%%%%%%%%%%%%%%%%%%%
\newsection{Introduction: asymptotic safety and dimensional reduction} 

Roughly speaking asymptotic safety is a property of a non-renormalizable 
quantum field theory replacing asymptotic freedom for a renormalizable one
\cite{Weinberg}. Unsurprisingly not many examples are known so far. 
Weinberg's original idea was that 4D quantum Einstein gravity might 
have this property in a suitable (non-perturbative) formulation, 
a scenario for which only recently non-trivial evidence has been reported 
\cite{Reuter02a,Reuter02b,Souma99,FalkOdi98}. Nevertheless it is unlikely 
that a definite conclusion can be reached on the issue in the near 
future. In view of this it seems worthwhile to look for simpler 
systems where a definite conclusion can be reached. For dimensionally
reduced gravity theories this turns out to be the case. 

Generalizing the results for pure gravity in \cite{PTletter,PTpaper} 
we consider here the 2-Killing vector reduction of 4D Einstein gravity 
coupled to abelian gauge fields and scalars in a way as they arise 
from $D\!\geq \!4$ dimensional (super-)~gravity theories \cite{BGM88}. 
Coupling gravity to scalars and/or gauge fields is known to destroy 
even one loop renormalizability \cite{tHVelt74,DesNieuw74} so that the 
status of the asymptotic safety scenario is of particular interest; 
see \cite{PercPern02} for some indicative results. We consider the 2-Killing   
vector reduction of the large class of 4D matter coupled systems 
studied in \cite{BGM88}; see also \cite{CJLP,Keu02} for their higher 
dimensional origin. For our purposes this class of matter couplings 
is natural because in their 2 Killing vector reduction all but two of 
the 4D gravitational + 4D matter degrees of freedom can be arranged 
to parameterize a non-compact symmetric space $G/H$. The reduced 
action becomes that of 2D gravity non-minimally coupled to a $G/H$ 
sigma-model, where the coupling is via the ``area radius'' $\rho$ of 
the two Killing vectors. This and 
the residual conformal factor $\sigma$ drastically change the classical 
and quantum dynamics as compared to the same noncompact $G/H$ sigma-model 
without coupling to gravity \cite{ZJbook}. The qualitative differences 
are summarized in the table below.  
%%%%%%%%%%%%%%%%%%%%%%%%%%%%%%%%%%
\vspace{5mm}

\hspace{2.3cm}
\begin{tabular}{|c|c|}
\hline
$G/H$ sigma-model            & dim.~red.~gravity with $G/H$   \\[0.5ex] 
\hline \hline
renormalizable               & non-renormalizable             \\
one essential coupling       & $\infty$ essential couplings   \\
$\lb$                        & function $h(\,\cdot\,)$        \\[1mm]
\hline
flow is  formally infrared free & flow is asymptotically safe \\[1mm]
\hline
trivial fixed point          & non-trivial fixed point        \\
$\lb=0$    &  $h^{\rm beta}(\,\cdot\,)$                       \\
 formally IR stable          & UV stable                      \\
 trace anomalous & trace anomaly vanishes                     \\
\hline           
\end{tabular}
%%%%%%%%%%%%%%%%%%%%%%%%%%%%%%%%%%%
\vspace{8mm}

The result is surprising for several reasons. First because the 
stability properties of the renormalization flow are reversed,
for which there is no obvious 2D reason. Second, a non-trivial 
fixed point exists in all cases, where gravity remains self-interacting 
and coupled to matter. The trace anomaly then vanishes 
and the quantum constraints (stemming from the residual 2D diffeomorphism 
invariance) can in principle be imposed. Third, although the 
structure of the symmetric space $G/H$ depends on the signature 
of the Killing vectors and the details of the reduction procedure,
the generalized beta function governing the functional $h$-flow is 
independent of it. These features are not built into the renormalization 
procedure as is highlighted by the fact that when 
a collection of 4D free massless fields is minimally coupled to 
4D Einstein gravity its 2-Killing vector reduction only has a 
trivial fixed point where the scalars decouple from gravity.
The latter is a special case of `augmented' scalar matter 
which we also consider.

The article is organized as follows. After introducing the class of 
4D gravity theories considered along with their 2-Killing vector reduction,
the renormalization of the basic Lagrangian is performed in section 3
to all loop orders and the generalized beta function for the essential
couplings is derived. The existence of an UV stable fixed point for 
their renormalization flow is established in section 4 and is extended
to additional scalar matter in section 5. In all cases the 
vanishing of the beta function(s) is subsequently found to entail
the vanishing of the trace anomaly modulo an improvement term, and vice 
versa. Some technical material is relegated to appendices.

%%%%%%%%%%%%%%%%%%%%%%%%%%%%%%%%%%%%%%%%%%%%%%%%%%%%%%%%%%%%%%%%%%%%%%%%
\newsection{Einstein gravity coupled to abelian gauge fields and scalars} 

Here we describe the class of 4D gravity theories considered and 
outline the reduction procedure. We largely follow the treatment 
in \cite{BGM88,BM00}. The higher dimensional origin of their 3D reductions
is systematically explored in \cite{CJLP,Keu02}. 

We consider 4D Einstein gravity coupled to $k$ abelian gauge fields 
and $\nbar$ scalars through the following action
\be
S_4 = \int d^4 x \sqrt{-g}\left[ - R(g) + \frac{1}{2} 
\bra \Jbar^{\alpha},\Jbar_{\alpha}\ket_{\gsfbar} - 
\frac{q}{4} F^T_{\alpha \beta} (\mu \,F^{\alpha\beta} - \nu\, 
{}^*\!F^{\alpha\beta})\right]\,.
\label{4Daction}
\ee  
Here $g_{\alpha\beta}\,,1 \leq \alpha,\beta \leq 4$, is the spacetime
metric with eigenvalues $(+,-,-,-)$, $R(g)$ is its scalar curvature
and indices are raised with $g^{\alpha\beta}$. There are $k$ real 
abelian vector fields arranged in a column 
$B_{\alpha}=(B_{\alpha}^I)$, $I = 1,\ldots ,k$, 
with field strength $F_{\alpha\beta} = \dd_{\alpha} B_{\beta} - 
\dd_{\beta} B_{\alpha}$ and dual field strength 
${}^*\!F^{\alpha\beta} = \frac{1}{2 \sqrt{-g}} 
\epsilon^{\alpha\beta\gamma\delta} F_{\gamma\delta}$.  
The scalars $\vpbar^i,\, i=1,\ldots, \nbar$, parameterize 
a non-compact riemannian symmetric space $\Gbar/\Hbar$ with 
metric $\mfb_{ij}(\vpbar)$. Its $\dim \Gbar$ Killing vectors
give rise to a Lie algebra valued Noether current $\Jbar_{\alpha}$ 
In terms of them the sigma-model Lagrangian for the scalars 
can be written as $\bra \Jbar^{\alpha}, \Jbar_{\alpha} \ket_{\gsfbar}$,
where $\bra\,\cdot\,,\,\cdot\,\ket_{\gsfbar}$ is an invariant scalar product 
on the Lie algebra $\gsfbar$. Finally the coupling matrices 
$\mu = \mu(\vpbar)$ and $\nu = \nu(\vpbar)$ are symmetric 
$k \times k$ matrices that depend on the scalars; the constant 
$q>0$ has been extracted for normalization purposes. The 
vector fields are supposed to contribute positively to the 
energy density which requires that $\mu$ is a positive definite
matrix. As such it has a unique positive square root $\mu^{1/2}$ 
to be used later. 
The coupling matrices $\mu$ and $\nu$ are now chosen in a way 
that renders the field equations derived from $S_4$ -- 
though in general not the action itself -- $\Gbar$-invariant.

The field equation for the gauge fields $\nabla_{\alpha}( \mu
F^{\alpha\beta} - \nu\,{}^*\!F^{\alpha\beta})=0$ can be interpreted  
as the Bianchi identity for a field strength $G_{\alpha\beta}= 
\dd_{\alpha}C_{\beta} - \dd_{\beta}C_{\alpha}$ derived from 
dual potentials $C_{\alpha}$. For later convenience one 
chooses ${}^*\!G_{\alpha\beta} = \eta(\mu F_{\alpha\beta} - \nu 
{}^*\!F_{\alpha\beta})$ with some constant orthogonal matrix $\eta$. 
In view of ${}^{**}\!F = - F$ they satisfy the linear relation
\be
{F \choose G} 
= \Upsilon \cV_{\csf} \cV_{\csf}^T {{}^*\!F \choose {}^*\!G}
\quad \mbox{with} \quad \cV_{\csf} = 
\left(\begin{array}{cc} \mu^{1/2} & \nu \mu^{-1/2} \\ 
0 & \eta \mu^{-1/2} \end{array} \right) \,,
\quad 
\Upsilon = 
\left(\begin{array}{cc} 0 & \eta^T \\ 
-\eta & 0\end{array} \right) \,,
\label{FGrel} 
\ee
where the subscript $\csf$ is mnemonic for `coupling'. 
If one now assumes that the column ${ F\choose G}$ transforms
linearly under a faithful 2k dimensional real matrix representation $\csf$  
of $\Gbar$, i.e. ${ F\choose G} \mapsto \csf(\bar{g}\inv)^T { F\choose G}$,
$\bar{g} \in \Gbar$, one finds that (\ref{FGrel}) transforms covariantly 
if $\cV_{\csf} \mapsto \csf(\bar{g}) \cV_{\csf} h_{\csf}$, with an 
orthogonal matrix $h_{\csf}$ and 
$\csf(\bar{g}^{-1})^T = \Upsilon \csf(\bar{g}) \Upsilon\inv$.    
Comparing this with the transformation law of the $\Gbar$-valued coset 
representatives $\cV_{\!*}$ in appendix A one sees that these 
conditions are satisfied if $\csf(\cV_{\!*}) = \cV_{\csf}$ $(*)$ and 
$\csf(\bar{\tau}(\bar{g})) = \csf(\bar{g}\inv)^T$, $\bar{g} \in \Gbar$,
where $\bar{\tau}$ is the involution whose set of fixed points defines
$\Hbar$. Clearly this restricts the allowed cosets $\Gbar/\Hbar$. For
the admissible ones Eq.~$(*)$ then determines the couplings 
$\mu(\vpbar)$, $\nu(\vpbar)$ as functions of the scalars. 
Since $\csf$ is faithful the determination is unique for a 
given choice of section $\cV_{\!*}$. Since $\cV_{\csf} 
\cV_{\csf}^T = \csf( \cV_{\!*} \bar{\tau} (\cV_{\!*}^{-1}))$ 
the result does not dependent on the choice of section, 
i.e.~$\widetilde{\cV}_{\!*} = \cV_{\!*} h$ for some $H$-valued
function $h$ determines the same $\mu(\vpbar)$ and $\nu(\vpbar)$.

Under the above conditions the 1-Killing vector reduction of $S_4$ 
can be brought into a form which -- after a partial dualization 
-- is manifestly $\Gbar$-invariant (while in the 4D theory $\Gbar$ 
was only an on-shell symmetry). By definition the fields entering 
the action are assumed to have vanishing Lie derivatives 
$\cL_K g_{\alpha\beta} = \cL_K B_{\alpha} = \cL_K \vpbar =0$, 
where $K^{\alpha}$ is the Killing vector and for the vector fields
a gauge has been chosen. The Killing vector can be spacelike
($\eps_1 = +1$) or timelike ($\eps_1 = -1$). As indicated we 
distinguish both cases by a sign and write $K^{\alpha}K_{\alpha} =: 
- \eps_1 \Delta$, with $\Delta >0$. Any tensor can then decomposed 
into components parallel and orthogonal to $K^{\alpha}$;
the respective projectors are $- \eps_1 \Delta^{-1} K_{\alpha} K^{\beta}$
and $\delta_{\alpha}^{\beta} + \eps_1 \Delta^{-1} K_{\alpha} K^{\beta}$. 
In particular the components of the metric orthogonal and parallel
to $K^{\alpha}$ are $g_{\alpha\beta} + \eps_1 \Delta^{-1} K_{\alpha} 
K_{\beta} =: \eps_1 \Delta \gamma_{\alpha\beta}$ and $K_{\alpha} 
=: - \eps_1 \Delta k_{\alpha}$, where we extracted $\Delta$ for 
later convenience. 
To solve the Killing equations one chooses adapted coordinates
in which $K = K^{\alpha} \dd_{\alpha}$ acts by translations; ordering
coordinates according to the assumed $(+,-,-,-)$ eigenvalues of 
the metric we take $K= \dd_3$ for $\eps_1 =1$ and $K= \dd_0$ for 
$\eps_1 =-1$. The fields then only depend on the remaining coordinates
which we rename $x^{\hat{\alpha}}, \hat{\alpha}=0,1,2$. In these 
coordinates we write $\eps_1 \Delta \gamma_{\hat{\alpha}\hat{\beta}}$ 
for the metric components orthogonal to $K^{\alpha}$ and 
$-\eps_1 \Delta k_{\hat{\alpha}}$ for those parallel to it.
Note that $\gamma_{\hat{\alpha}\hat{\beta}}$ has eigenvalues 
$(+,-,-)$ for $\eps_1 = +1$ and $(+,+,+)$ for $\eps_1 =- 1$. For 
the vector fields one then has $B_{\hat{\alpha}} = k_{\hat{\alpha}} B 
+ B_{\hat{\alpha}}^{\perp}$, with $B = B_{\alpha}K^{\alpha}$ and 
$B_{\alpha}^{\perp} K^{\alpha} =0$.     

Entering with this decomposition into the action $S_4$ yields a
reduced action $S_3^{\rm direct}$, where a 3D Einstein-Hilbert term
for the metric $\gamma_{\hat{\alpha}\hat{\beta}}$ is minimally 
coupled to the 3D `matter' fields $\Delta,k_{\hat{\alpha}}$ and 
$\vpbar,\,B,\,B_{\hat{\alpha}}^{\perp}$. On general grounds (the 
``principle of symmetric criticality'', see e.g.~\cite{TorreFels01}) 
the field equations derived from $S_3^{\rm direct}$ coincide with 
the reduction of the original field equations. In particular the ones
for $k_{\hat{\alpha}}$ and $B_{\hat{\alpha}}^{\perp}$ can be 
interpreted as Bianchi identities
implying the existence of scalar potentials $\psi$ and $C = (C^I),\,
I =1,\ldots, k$, where the latter are just the parallel components
of the 4D dual vector potentials, $C = C_{\alpha} K^{\alpha}$. 
The defining relations are 
\ba
&\nspace & \dd^{\hat{\alpha}} k^{\hat{\beta}} - 
\dd^{\hat{\beta}} k^{\hat{\alpha}} = \frac{1}{\sqrt{\gamma}} 
\eps^{\hat{\alpha}\hat{\beta}\hat{\gamma}} \frac{1}{\Delta^2} 
\omega_{\hat{\gamma}} \quad \mbox{with} \quad 
\omega_{\hat{\alpha}} = \dd_{\hat{\alpha}} \psi + 
\frac{q}{2}[C^T \eta \dd_{\hat{\alpha}} B - B^T 
\eta^T \dd_{\hat{\alpha}} C]\,,
\nonum 
&\nspace & \dd^{\hat{\alpha}} (B^{\perp})^{\hat{\beta}} -
\dd^{\hat{\beta}} (B^{\perp})^{\hat{\alpha}} + 
(\dd^{\hat{\alpha}} k^{\hat{\beta}} - \dd^{\hat{\beta}} k^{\hat{\alpha}}) B 
= - \frac{\eps_1}{\sqrt{\gamma}} \eps^{\hat{\alpha}\hat{\beta}\hat{\gamma}} 
\frac{1}{\Delta} \mu^{-1}
[\nu \dd_{\hat{\gamma}} B - \eta^T \dd_{\hat{\gamma}} C] \,.    
\label{3Dduality}
\ea
Performing a Legendre transformation to the scalar variables 
one obtains \cite{BGM88} 
\ba
\label{3Daction} 
&& S_3 = \int d^3 x \sqrt{\gamma}\left[ R^{(3)}(\gamma) -
\frac{1}{2} \gamma^{\hat{\alpha}\hat{\beta}} \mf_{ij}(\vp) 
\dd_{\hat{\alpha}} \vp^i \dd_{\hat{\beta}} \vp^j \right]\,, 
\quad \mbox{with} 
\\[2mm] 
&& \mf_{ij}(\vp)\dd_{\hat{\alpha}} \vp^i \dd_{\hat{\beta}} \vp^j = 
\frac{1}{\Delta^2}(\dd_{\hat{\alpha}}\Delta \dd_{\hat{\beta}} \Delta
+ \omega_{\hat{\alpha}}\omega_{\hat{\beta}})+  
\bra \Jbar_{\hat{\alpha}},\Jbar_{\hat{\beta}}\ket_{\gsfbar} 
+\eps_1 \frac{q}{\Delta} 
\dd_{\hat{\alpha}} A^T \cV_{\csf} \cV_{\csf}^T \dd_{\hat{\beta}} A\,.  
\nonumber
\ea
Here the scalars $B,C$ have been arranged into a 2k dimensional
column $A = {B \choose C}$ and since $\omega_{\hat{\alpha}} = 
\dd_{\hat{\alpha}} \psi - \frac{q}{2} A^T \Upsilon \dd_{\hat{\alpha}} A$ 
the action (\ref{3Daction}) is manifestly $\Gbar$-invariant. Moreover 
we anticipated that the 3D `matter' combines into a nonlinear sigma-model. 
The target space has dimension $2+ \nbar + 2 k$, we take 
$\vp^T = (\Delta,\psi,\vpbar^T, A^T)$ as field coordinates, and
with the normalization $\bra \Jbar_{\hat{\alpha}},
\Jbar_{\hat{\beta}}\ket_{\gsfbar} = \mfb_{ij}(\vpbar)
\dd_{\hat{\alpha}} \vpbar^i \dd_{\hat{\beta}}\vpbar^j$ the metric 
comes out as
\be
\mf(\vp) = \left(\begin{array}{cc|c|c}
\frac{1}{\Delta^2} & 0 &  \mbox{} &  0 \\
0 & \frac{1}{\Delta^2} &  \mbox{} & -\frac{q}{2 \Delta^2} A^T \Upsilon 
\\[1mm] \hline & & & \\[-2mm]
\mbox{}  & \mbox{} & \quad \mfb(\vpbar) \quad & \mbox{} \\[3mm]
\hline
\\[-6mm] \mbox{} & \mbox{} & \mbox{} \\[-5mm]
0  & \frac{q}{2\Delta^2} \Upsilon A & \mbox{} & 
\eps_1\frac{q}{\Delta} \cV_{\csf}\cV_{\csf}^T - \frac{q^2}{4 \Delta^2} 
\Upsilon A \otimes A^T \Upsilon 
\end{array}
\right).
\label{mmetric}
\ee
It is riemannian for $\eps_1 = +1$ and pseudo-riemannian for $\eps_1 =-1$,
with $2k$ negative eigenvalues. Note also that due to the dualization 
the part of the coset space parameterized by the gravitational 
potentials $(\Delta,\psi)$ always has eigenvalues $(++)$.

We briefly digress on the isometries of (\ref{mmetric}). By virtue 
of the $\Gbar$ invariance of the action $\mf$ has $\dim \Gbar$ Killing 
vectors of which $\nbar = \dim \Gbar/\Hbar$ are algebraically independent. 
The residual gauge transformations $A \mapsto A + a$, $\psi \mapsto 
\psi - \frac{q}{2}A^T \Upsilon a$, with a constant $2k$ column $a$ 
give rise to $2k$ Killing vectors. Finally constant translations in 
$\psi$ and scale transformations $(\Delta,\psi,\vpbar^T, A^T) \mapsto 
(\lb \Delta, \lb \psi, \vpbar^T, \lb^{1/2} A^T)\,,\lb >0$, 
are obvious symmetries of the action. The associated Killing vectors
${\bf e},{\bf h}$ of $\mf$ generate a Borel subalgebra of $sl_2$, i.e.
$[{\bf h},{\bf e}] = - 2 {\bf e}$. In contrast the last $sl_2$ generator 
${\bf f}$ is only a Killing vector of $\mf$ under certain conditions 
on $\Gbar/\Hbar$. If these are satisfied a remarkable `symmetry 
enhancement' takes place in that $\mf$ is the metric of a 
much larger symmetric space $G/H$, where $G$ is a non-compact real 
form of a simple Lie group with $\dim G = \dim \Gbar + 
4k + \dim SL(2)$. The point is that if ${\bf f}$ exists as Killing 
vector its commutator with the gauge transformations is nontrivial 
and yields $2k$ additional symmetries (generalized 
``Harrison transformations''). Since $\mf$ always has $\dim \Gbar + 2 k +2$
Killing vectors the additional $1 + 2k$ then match the dimension of $G$. 
For the number of dependent Killing vectors, i.e.~the dimension 
of the putative maximal subgroup $H \subset G$ one  expects 
$\dim H = \dim \Hbar + 1 + 2 k$. Indeed under the conditions stated
the symmetric space $G/H$ exists and is uniquely determined by 
$\Gbar/\Hbar$ and the sign $\eps_1$. See \cite{BGM88} for a complete list. 
Evidently the gauge fields are crucial for the symmetry enhancement.
among the systems in \cite{BGM88} only pure gravity has $k=0$.

We proceed by performing the reduction with respect to the second
Killing vector. Clearly if the first Killing vector was timelike 
it can only be spacelike $(\eps_2 =+1)$ while if the first Killing
vector was spacelike the second can be either spacelike $(\eps_2 =+1)$ 
or timelike $(\eps_2 =-1)$. Again one chooses adapted coordinates in 
which the second Killing vector acts by translations. The fields in $S_3$ 
will then only depend on the remaining two non-Killing coordinates
which we denote by $x^{\mu}, \mu =0,1$. The decomposition of the 
3D metric $\gamma_{\hat{\alpha}\hat{\beta}}$ can be performed as before. 
Anticipating that the off-diagonal components turn out to be non-dynamical 
we write $-\eps_1\eps_2 \rho^2$ for the component contracted with 
the Killing vector and $\gamma_{\mu\nu}$ for the block orthogonal to it, 
which is diffeomorphic to ${\rm diag}(\eps_2 e^{\sigma},-\eps_1 e^{\sigma})$.  
One readily checks that the sign pattern matches the $(+, -\eps_1,-\eps_1)$ 
eigenvalues of the 3D metric. Entering with this decomposition into 
$S_3$ yields the 2D action $S_2 = \int \!d^2 x \,L_{h=\rho}$, which we 
shall take as the starting point to develop the quantum theory. 
We write the Lagrangian as a special case of a family of 
Lagrangians $L_h$ labeled by a real function $h$ of one 
variable and two parameters $a,b\in \R,\,b\neq 0$, because this is what 
is needed in the quantum theory. The generalized Lagrangian reads 
\be
L_h(\vp,\rho,\sigma) = \frac{1}{2\lb} \,h(\rho) \sqrt{\gamma}  
\gamma^{\mu\nu} \Big[\mf_{ij}(\vp) \dd_{\mu} \vp^i \dd_{\nu} \vp^j 
+ a \rho^{-2} \dd_{\mu} \rho \dd_{\nu} \rho \Big] + \frac{1}{2\lb} 
f(\rho) \sqrt{\gamma} R^{(2)}(\gamma)\,.
\label{act1}
\ee 
Here $\lb$ is Newton's constant per unit volume of the internal space. 
The 2D metric $\gamma_{\mu\nu}$ is diffeomorphic to 
$\eta_{\mu\nu} e^{\sigma}$, where $\eta_{\mu\nu}$ is constant with
eigenvalues $(\eps_2, -\eps_1)$. Its scalar curvature 
$R^{(2)}(\gamma)$ is normalized such that $R^{(2)}(e^{\sigma} \eta) 
= - e^{-\sigma} \dd^2 \sigma$. The target space metric is that 
of Eq.~(\ref{mmetric}); note that its signature only depends on $\eps_1$, 
i.e.~on the spacelike/timelike character of the first Killing vector. 
In the quantum theory the function $h$ parameterizes an infinite set 
of essential couplings (e.g.~via the expansion coefficients with respect 
to some basis) while classically $h(\rho) = \rho$. The function 
$f$ is given in terms of $h$ by  
\begin{equation}
f(\rho) = 2 b \int^{\rho} \frac{du}{u} h(u) \,,
\label{f} 
\end{equation}
for some non-zero constant $b$. Likewise $a$ is a real constant;
the values $b=-1, a=0$ for $h(\rho) =\rho$ are the ones that come out 
of the reduction procedure. 

Since $\rho$ will play a pivotal role in the following it is worth 
pointing out its geometrical meaning as the `area radius' associated 
with the two Killing vectors. Let $\kappa_1,\,\kappa_2$ and $\kappa_{12}$ 
denote the norm of the two Killing vectors and their inner product, 
respectively, with respect to the 4D metric $g_{\alpha\beta}$. 
By definition $-\eps_1 \kappa_1^2 = \Delta \geq 0$ and 
$- \eps_2 \kappa_2^2 \geq 0$. Spelling out the net parameterization 
of $g_{\alpha\beta}$ induced by the above procedure (but without 
dualizing the $k_{\hat{\alpha}}$ via (\ref{3Dduality})) one finds 
\be
\rho^2 = \eps_1 \eps_2 \big(\kappa_1^2 \kappa_2^2 - 
\kappa_{12}^2 \big) \geq 0\,,
\sspace 
g^{\alpha \beta} \dd_{\alpha} \rho \dd_{\beta} \rho 
= \eps_1 \Delta e^{-\sigma} \dd^{\mu} \rho \dd_{\mu} \rho\,. 
\label{rhoKillings}
\ee
If one of the Killing vectors has compact orbits of length $2\pi$, 
the area swept out per unit length of the other is $2\pi \rho$. 
Whence the term ``area radius'' \cite{Goncalves} which we retain 
also in the general case. The gradient of $\rho$ is spacelike with 
respect to the 4D metric if one of the Killing vectors is timelike 
and indefinite if both are spacelike.

The action (\ref{act1}) is manifestly invariant under 
2D diffeomorphisms. The associated hamiltonian and 1D diffeomorphism 
constraints are $\cH_0 = T_{00}$ and $\cH_1 = T_{01}$, where 
$T_{\mu\nu}$ is the energy momentum tensor of the (flat space) 
action (\ref{act1}) in conformal gauge 
$\gamma_{\mu\nu} = e^{\sigma} \eta_{\mu\nu}$. As usual light cone 
components $T_{\pm\pm} \sim \cH_0 \pm \cH_1$ are convenient  
\begin{equation}
\lb T_{\pm\pm} = h(\rho)[\mf_{ij}(\vp)\dd_{\pm} \vp^i \dd_{\pm} \vp^j 
+ a \rho^{-2} (\dd_{\pm} \rho)^2] + 
\dd_{\pm} \sigma \dd_{\pm} f - \dd_{\pm}^2 f\;.
\label{constr}
\end{equation}
They constitute a pair of first class constraints and generate two 
commuting copies of a centerless Virasoro-Witt algebra with respect to 
the Poisson structure induced by (\ref{act1}). Importantly the latter 
coincides with the symplectic structure induced from the higher 
dimensional theory.   

Even in conformal gauge the classical action (\ref{act1}) `remembers' its 
gravitational origin. Recall that in a diffeomorphism invariant theory 
the Lagrangian can always be written as a total divergence on-shell.
For the action (\ref{act1}) in conformal gauge there exist two currents 
$C_{\mu},\,D_{\mu}$ obeying  
\ba
\nspace \dd^{\mu} C_{\mu} \is \rho \dd_{\rho} \ln h \cdot L_h\;,
\quad \;\sspace \; \;C_{\mu} = 
\frac{b}{2 \lb}h(\rho) \dd_{\mu}(2 \sigma + \frac{a}{b} \ln \rho)\,,
\nonum
\nspace \dd^{\mu} D_{\mu} \is -\ln \rho \cdot \rho \dd_{\rho} \ln h 
\cdot L_h\;,\quad\, D_{\mu} = \frac{b}{\lb}h(\rho)
(\sigma \dd_{\mu} \ln \rho - \ln \rho \dd_{\mu} \sigma)\,. 
\label{cward1}
\ea
In particular there are (exactly) two choices for $h(\rho)$ for which the 
Lagrangian is a total divergence on-shell: $h(\rho) \sim \rho^p,p\neq 0$, 
and $h(\rho) \sim \ln\rho$. The first one corresponds to the outcome
of the classical reduction procedure (and, as will become clear below,
it is also necessary and sufficient for one-loop renormalizability). 

In summary, starting from 4D gravity coupled to vectors and scalars 
according to (\ref{4Daction}) we performed a two step 
dimensional reduction procedure. In the first step only one Killing 
vector is used and in the resulting 3D theory all vectors are replaced 
by dual scalar potentials. The original and the induced scalars 
combine under certain conditions to parameterize a symmetric space 
of the form $G/H$, where $G$ is a non-compact real form of a simple 
Lie group and $H$ is a maximal subgroup (whose properties depend on 
the signature of the Killing vector). The resulting 3D action $S_3$ 
couples 3D gravity minimally to a $G/H$ sigma-model. Finally the 
reduction with respect to the second Killing vector is performed. 
This leaves the coset $G/H$ unaffected while the 3D metric gives rise 
to the scalars $\rho$ and $\sigma$. The classical action $S_2$ 
derives from the Lagrangian (\ref{act1}) with $h(\rho) =\rho$ 
and $b=-1,a=0$. The generalized form (\ref{act1}) anticipates
the structure of the renormalized Lagrangian.

%%%%%%%%%%%%%%%%%%%%%%%%%%%%%%%%%%%%%%%%%%%%%%%%%%%%%%%%%%%%%%%%%%%%%%%%%
\newpage
\newsection{Universal dressing of the beta function}

The goal in the following is to construct a perturbative quantum 
theory based on the Lagrangian (\ref{act1}). We fix the conformal 
gauge for $\gamma_{\mu\nu}$ and aim at a Dirac quantization. 
Further the renormalization strategy of \cite{PTpaper} will be 
adopted, in particular we use the covariant background 
field expansion, dimensional regularization, and minimal subtraction to
determine the counter terms. For the cancellation of the counter 
terms nonlinear field renormalizations are allowed, nevertheless 
one finds that the system {\it cannot} be renormalized
with {\it finitely} many essential couplings beyond one loop. 
Hence an {\it infinite} set of essential couplings is required,
identifying the systems as non-renormalizable ones. 
The systems remain managable because these couplings can be combined 
into (the expansion coefficients of) one function $h$ of $\rho$ which 
enters the renormalized Lagrangian in the way anticipated in (\ref{act1}). 
The fact that $h$ indeed qualifies as an ``essential'' set of couplings
can be seen from the identities (\ref{cward1}). Beyond one loop one
is forced to take $h(\rho)$ different from $\rho^p$ upon variation 
of the renormalization scale. Then $\dd^{\mu} C_{\mu} \not\sim L_h$ 
and the expansion coefficients of $h$ with respect to some basis are 
essential couplings. Though we shall only be interested in the systems 
(\ref{act1}) arising by dimensional reduction, the results of section 
3.1 and 3.2 hold for {\it any} symmetric space $G/H$ with $G$ simple 
(i.e.~not only for the specific non-compact ones in (\ref{mmetric})) 
as well as for direct products thereof with identical factors. The 
non-compactness of $G/H$ will be crucial in section 4.

%%%%%%%%%%%%%%%%%%%%%%%%%%%%%%%%%%%%%%%%%%%%%%%%%%%%%%%%%%%%%%%%%%%%
\newsubsection{Renormalization of the Lagrangian to all loop orders}

In order to determine the counter terms we interpret (\ref{act1}) as the 
action of a riemannian sigma-model in the sense of Friedan \cite{Frie85}. 
Introducing two coordinates $\phi^{n+1} := \rho,\,\phi^{n+2} 
:= \sigma$  in addition to $\phi^i = \vp^i, \,i =1,\ldots,n$, 
one can regard (\ref{act1}) as a sigma-model with a fiducial 
non-homogeneous target manifold of dimension $n+2$. In conformal 
gauge $\gamma_{\mu\nu} = e^{\sigma} \eta_{\mu\nu}$ and after integration 
by parts one has $L_h(\vp,\rho,\sigma) = \frac{1}{2\lb} 
\gf_{ij}(\phi) \dd \phi^i \dd \phi^j$, with 
\be
\gf_{ij} = h(\rho) 
\left( \begin{array}{cc|c}
              {}               & {} & {}\\[-3mm] 
(\mf_{ij})_{1\leq i,j\leq n} & {} & \mbox{\LARGE 0}\\[-2mm]
              {}               & {} & {} \\[-1mm]  
\hline 
\mbox{\LARGE 0} & {} & \begin{array}{cc} a/\rho^2 & b/\rho \\
                                         b/\rho & 0 \end{array}
\end{array} \right)\,.
\label{Tmetric}
\ee
Here $\mf_{ij}$ is the metric (\ref{mmetric}); as remarked above 
the results of this and the following section remain valid if 
it is replaced by the metric of an arbitrary symmetric space 
$G/H$ with $G$ simple. In addition to the Killing vectors associated 
with $\mf$ the metric (\ref{Tmetric}) possesses two conformal Killing 
vectors ${\bf t}_+ =
\rho \dd_{\rho} - \frac{a}{2b} \dd_{\sigma}$ and ${\bf d} = 
- \rho \ln \rho \dd_{\rho} +(\sigma + \frac{a}{b} \ln \rho) \dd_{\sigma}$,  
which together with ${\bf t}_- = \dd_{\sigma}$  
generate the isometries of $\R^{1,1}$. The currents associated with 
${\bf t}_+$ and ${\bf d}$ are those in Eq.~(\ref{cward1}).

In dimensional regularization $(\int\! d^2 x \rra \int \!d^dx)$ 
the $l$-loop counter terms contain 
poles of order $\nu \leq l$ in $(2-d)$. We denote the coefficient 
of the $\nu$-th order pole by $T^{(\nu,l)}_{ij}(\gf)$. In principle
the higher order pole terms are determined recursively by the
residues $T^{(1,l)}_{ij}(\gf)$ of the first order poles. Taking the 
consistency of the cancellations for granted one can focus on 
the residues of the first oder poles, which we shall do throughout. 
In appendix B we show that to all loop orders they have 
the following structure: 
\be
T^{(1,l)}_{ij}(\gf) = \frac{1}{h(\rho)^{l-1}} 
\left( \begin{array}{cc|c}
             {}               & {} & {}\\[-2mm] 
\zeta_l\; (\mf_{ij})_{1\leq i,j\leq n} & {} & \mbox{\LARGE 0}\\[-2mm]
            {}               & {} & {} \\  
\hline 
\mbox{\LARGE 0} & {} & \begin{array}{cc} 
                            \frac{n}{\rho^2} S_l(\rho) & 0\\
                                                      0  & 0 \end{array}
\end{array} \right)\,,\sspace \forall\, l \geq 1\,.
\label{Tcounter}
\ee
The $\zeta_l$ are constants defined through the curvature scalars 
of $\mf_{ij}$. The $S_l(\rho)$ are differential polynomials in $h$ 
invariant under constant rescalings of $h$ and normalized to vanish 
for constant $h$. The first three are:
\ba
S_1(h) \is -\frac{1}{2}(\rho \dd_{\rho})^2 \ln h + 
\frac{1}{4} (\rho \dd_{\rho} \ln h)^2\,,\sspace 
S_2(h) =0\,,
\nonum
S_3(h) \is -\frac{\zeta_2}{4}(\rho \dd_{\rho})^2 \ln h + 
\frac{\zeta_2}{12} (\rho \dd_{\rho} \ln h)^2\,.
\label{Sl}
\ea
The counter terms (\ref{Tcounter}) ought to be absorbed by nonlinear 
field renormalizations 
\be
\phi^j_{\b} = \phi^j + \frac{1}{2-d} \Xi^j(\phi,\lb) + \ldots\,,
\quad \mbox{with} \quad \Xi^j = \sum_{l \geq 1} 
\Big(\frac{\lb}{2\pi}\Big)^l \phi_l^j(\phi)\,, 
\label{phiren}
\ee
and a renormalization of the function $h$
\be 
h_{\!\b}(\rho) = \mu^{d-2} h(\rho,\lb) \left[
1 + \frac{1}{2-d} H(\rho,\lb) + \ldots \right] \,,
\quad H(\rho,\lb) =  \sum_{l \geq 1} \Big(\frac{\lb}{2\pi}\Big)^l 
H_l(\rho)\;,
\label{hren}
\ee
where $\mu$ is the renormalization scale. Note that on {\it both} sides 
of (\ref{hren}) the argument is the renormalized field. The 
renormalized $h$ function is allowed to depend on $\lb$; specifically 
we assume it to have the form 
\be
h(\rho,\lb) = \rho^p + \frac{\lb}{2\pi} h_1(\rho) + 
\Big(\frac{\lb}{2\pi} \Big)^2 h_2(\rho) + \ldots \,, 
\label{hlb} 
\ee
where the first term ensures standard renormalizability at the 1-loop 
level -- and is determined by this requirement up the power $p\neq 0$. 
The power has no intrinsic significance; one could have chosen a 
parameterization of the 4D spacetime metric $g_{\alpha\beta}$ 
such that $h(\rho)= \rho^p$ in (\ref{act1}) was the outcome
of the classical reduction procedure. In particular the sectors 
$p> 0$ and $p<0$ are equivalent and we assume $p>0$ throughout.

Combining (\ref{Tmetric}), (\ref{phiren}), (\ref{hren}) and 
(\ref{Tcounter}) one finds that the first order poles cancel 
in the renormalized Lagrangian iff the following ``finiteness condition'' 
holds: 
\be
\cL_{\Xi}\, \gf_{ij} + H(\rho,\lb) \gf_{ij} = \lb T^{(1)}_{ij}(\gf/\lb) \,,
\label{finiteness}
\ee
where $T^{(1)}(\gf) = \sum_{l \geq 1} (\frac{1}{2\pi})^l 
T^{(1,l)}(\gf)$ and $\cL_{\Xi}\gf$ is a Lie derivative. The 
$\rho$-dependence of $H$ marks the deviation from conventional
renormalizability. Guided by 
the structure of (\ref{Tmetric}) 
and (\ref{Tcounter}) we search for a solution with $\Xi^j = 
(0 ,\ldots, 0, \Xi^{\rho}(\rho,\lb), \Xi^{\sigma}(\rho,\lb))$, where
here and later on we also use $\rho = n+1$, $\sigma = n+2$ for the 
index labeling. The Lie derivative term with this $\Xi^j$ is 
\ba
&& \cL_{\Xi}\, \gf_{ij} = 
\left( \begin{array}{cc|c}
              {}               & {} & {}\\[-2mm] 
\Xi^{\rho}(\rho)\dd_{\rho} h \; (\mf_{ij})_{1\leq i,j\leq n} 
                                   & {} & \mbox{\LARGE 0}\\[-2mm]
              {}               & {} & {} \\  
\hline 
\mbox{\LARGE 0} & {} & 
\begin{array}{cc} \cL_{\Xi}\gf_{\rho\rho}  & \cL_{\Xi}\gf_{\rho\sigma}\\
                   \cL_{\Xi}\gf_{\rho\sigma} & 0 \end{array}
\end{array} \right)\,,\\[4mm] \nonumber
&& \cL_{\Xi}\,\gf_{\rho\rho} 
= \frac{a}{\rho^2}\left[ \dd_{\rho} h \,\Xi^{\rho} + 
2 h \rho \dd_{\rho}\Big( \frac{\Xi^{\rho}}{\rho} \Big) \right]
+ 2 b \,\frac{h}{\rho} \,\dd_{\rho} \Xi^{\sigma}\,,
\quad 
\cL_{\Xi}\,\gf_{\rho\sigma} 
= b \,\dd_{\rho} \Big( \frac{h \Xi^{\rho}}{\rho} \Big)\,.
\label{Lieg}
\ea
The finiteness condition (\ref{finiteness}) then is equivalent to 
a simple system of differential equations whose solution is
%the 
%following system of differential equations:
%\ba 
%\label{findiffs}
%&& \Xi^{\rho}\, \dd_{\rho} h + H h = h B_{\lb}(\lb/h)\,, \nonum
%&& \rho \dd_{\rho} \left( \frac{h \Xi^{\rho}}{\rho} \right) + H h =0\,, 
%\\[2mm]
%&& \rho^2 X + a\, h H = n h S(\rho,\lb)\,,\nonumber
%\ea 
%The solution of (\ref{findiffs}) is: 
%where we set 
\ba
\label{finsol}
H(h/\lb) \is - \frac{1}{h(\rho,\lb)} \rho \dd_{\rho} 
\left[ h(\rho,\lb) \frac{\Xi^{\rho}(h/\lb)}{\rho} \right]\,,
\nonum
\Xi^{\rho}(h/\lb) \is -\rho \int^{\rho}\frac{du}{u} 
B_{\lb}\Big(\frac{\lb}{h(u,\lb)} \Big)\,,
\\[2mm]
\Xi^{\sigma}(h/\lb) \is - \frac{a}{2 b \, \rho} \Xi^{\rho}(h/\lb) 
+ \frac{1}{2b} \int^{\rho} \frac{du}{u} S(u,\lb) \,.
\nonumber
\ea
Here we set 
\be
B_{\lb}(\lb) := \sum_{l \geq 1} \zeta_l \Big(\frac{\lb}{2\pi}\Big)^l\,,
\sspace 
S(\rho,\lb) :=  n\sum_{l \geq 1} \Big(\frac{\lb}{2\pi}\Big)^l h^{-l} 
S_l(\rho) \,,
\label{BSdef}
\ee
and slightly adjusted the notation to stress the functional 
dependence on $h/\lb$. Possibly $\lb$-dependent integration constants 
have been absorbed into the lower integration boundaries of the integrals.
Throughout these solutions should be read as shorthands for 
their series expansions in $\lb$ with $h$ of the form 
(\ref{hlb}). For example 
\be
\Xi^{\rho}(\rho,\lb) = \frac{\lb}{2\pi} \frac{\zeta_1}{p} \rho^{-p+1}
+ \Big(\frac{\lb}{2\pi}\Big)^2 \rho \int_{\rho}^{\infty} 
\frac{du}{u^{2 p+1}} [\zeta_2 - \zeta_1 h_1(u)] + O(\lb^3)\,.
\label{Xiexpansion}
\ee

For the derivation of (\ref{finiteness}) and (\ref{finsol}) we 
fixed a coordinate system in which the 
target space metric takes the form (\ref{Tmetric}). Under a change 
of parameterization $\phi^j \ra \phi^j(\widehat{\phi})$ the 
finiteness condition (\ref{finiteness}) should transform covariantly,
and indeed it does. The constituents transform as 
\ba
\widehat{\gf}_{ij}(\phihat) \is 
\frac{\dd \phi^k}{\dd \phihat^i}    
\frac{\dd \phi^l}{\dd \phihat^j} \,\gf_{kl}(\phi) \,,
\sspace 
T^{(1)}_{ij}(\widehat{\gf})(\phihat) = 
\frac{\dd \phi^k}{\dd \phihat^i}    
\frac{\dd \phi^m}{\dd \phihat^j} T^{(1)}_{km}(\gf)(\phi) \,,
\nonum
\widehat{\Xi}^j(\phihat) \is \frac{\dd \phihat^j}{\dd \phi^k} 
\,\Xi^k(\phi)\,,
\sspace \quad\; \;
(\cL_{\widehat{\Xi}} \,\widehat{\gf})_{ij}(\phihat) = 
\frac{\dd \phi^k}{\dd \phihat^i}    
\frac{\dd \phi^m}{\dd \phihat^j} 
(\cL_{\Xi} \,\gf)_{km}(\phi)\,.
\label{fin_cov}
\ea    
The covariance of the counter terms as a function of the full field 
is nontrivial \cite{HPS88,BonnDeld86} and is one of the main advantages 
of the covariant background field expansion. The relations (\ref{fin_cov}) 
can be used to convert the solutions (\ref{finsol}) of the finiteness
condition into any desired coordinate system on the target space. 
The coordinates $\sigma$ and $\rho$ used in (\ref{Tmetric}) are adapted 
to the Killing vector ${\bf t}_-$ and the conformal Killing vectors
${\bf t}_+$, ${\bf d}$.

%%%%%%%%%%%%%%%%%%%%%%%%%%%%%%%%%%%%%%%%%%%%%%%%%%%%%%%%%%%%%%%%%%%%%%
\newsubsection{Universal dressing of the beta function} 

So far the renormalization has been performed at some fixed normalization
scale $\mu$. Changing the scale gives rise to renormalization
flow equations of which the one for the essential couplings $h$ is of 
primary interest. We first present the result and then outline the derivation. 
Denoting the `running' coupling function by $\hbbar(\,\cdot\,,\mu)$ 
the flow equation reads
\be
\mu \frac{d}{d\mu} \hbbar = \lb \betabf_h(\hbbar/\lb)\;.
\label{hflow}
\ee
The associated beta functional is given by  
\be
\lb \betabf_h(h/\lb) = - \rho \dd_{\rho} \left[ 
h \int_{\rho}^{\infty} \frac{du}{u} \frac{h(u)}{\lb} 
\beta_{\lb} \Big(\frac{\lb}{h(u)} \Big) \right]\,.
\label{betah} 
\ee
Here $\beta_{\lb}(\lb) = \sum_{l \geq 1} l \zeta_l (\frac{\lb}{2\pi})^l$ 
is the conventional beta function of the $G/H$ symmetric space 
sigma-model without coupling to gravity, defined e.g.~in the 
minimal subtraction scheme.  
Thus $\betabf_h(h)$ can be viewed as a ``gravitationally dressed'' version 
of $\beta_{\lb}(\lb)$, akin to the phenomenon in \cite{KlKoPo93,PSZ97}. 
Remarkably the ``dressing'' 
is universal, i.e.~independent of the symmetric space considered, 
and can be given in closed form to all loop orders. 

The flow equation (\ref{hflow}) of course has to be supplemented
by a boundary condition. The condition $h(\rho,\lb)/\rho^p \ra 1$ 
for $\rho \ra \infty$ entails that $\betabf_h(h)$ vanishes for 
$\rho \ra \infty$. This asymptotics is therefore preserved
by the flow. Moreover is can be seen to ensure that the flow is 
entirely driven by the counter terms, as it should. We therefore
adopt this asymptotic boundary condition throughout.

The derivation of (\ref{hflow}), (\ref{betah}) is straightforward
once $H$ is known explicitly as a functional of $h$ via (\ref{finsol}). 
Starting from (\ref{hren}) one determines how the renormalized 
$h(\,\cdot\,)$ (the function not its value) has to change upon a 
change of the renormalization scale in order to have the bare 
$h_{\b}(\,\cdot\,)$ scale independent. One finds
\be
\lb \betabf_h(h/\lb) = (2-d) h(\rho) - h(\rho) \int \!du \,h(u) 
\frac{\delta H(\rho,\lb)}{\delta h(u)}\,. 
\ee
Inserting (\ref{finsol}) and setting $d=2$ for simplicity yields the 
announced result (\ref{hflow}), (\ref{betah}).

%%%%%%%%%%%%%%%%%%%%%%%%%%%%%%%%%%%%%%%%%%%%%%%%%%%%%%%%%%%%%%%%%%%%%%%%
\newsubsection{Beta function coefficients and examples}

The coset spaces $G/H$ arising by dimensional reduction from 4 dimensions
fall into 5 infinite series associated with the `classical' groups and
10 isolated cases associated with exceptional groups. The leading beta
function coefficient $\zeta_1$ turns out to be given by the 
simple formula $\zeta_1 = - (k+2)/2$, where $k$ is the number of 
vector fields in (\ref{4Daction}). We postpone its derivation to 
section 4 where 
$\zeta_1$  will play a pivotal role. The next coefficient $\zeta_2$ 
is still scheme independent while the higher orders depend on the 
choice of the renormalization scheme; as described we adopt the 
minimal subtraction scheme throughout.  Explicit results for 
$\zeta_2$ and $\zeta_3$ can in principle be obtained by  
evaluating the expressions (\ref{T123}) in the vielbein frame. 
On account of (\ref{GHRiemann}) this reduces the computation to a group 
theoretical one. In contrast to $\zeta_1$ however details 
about the embedding $H\subset G$ enter, which is somewhat cumbersome 
to describe because $H$ is in general not simple even though $G$ is. 
It is therefore more convenient to
rely on the known results for the compact symmetric spaces
\cite{BrezHik78,Hikami81} and to translate them into the situation 
needed by using the fact that compact and non-compact symmetric spaces 
come in dual pairs \cite{Helgason}.  

In brief, a symmetric space $G_d/H_d$ is said to be dual to 
$G/H$ if the commutation relations in $\hsf$ are preserved but 
the bilinear form flips sign and vice versa for the $[\msf, \msf] \subset 
\hsf$ relations and the bilinear form on $\mf$. In the notation of 
appendix A 
\ba
&& {}_{d}f_{\dot{a}\dot{b}}^{\;\;\;\dot{c}} = 
\phantom{-} f_{\dot{a}\dot{b}}^{\;\;\;\dot{c}}\,,\sspace 
\bra t_{\dot{a}} ,t_{\dot{b}} \ket_{\gsf_d} = - 
\bra t_{\dot{a}} ,t_{\dot{b}} \ket_{\gsf}\,,
\nonum 
&& {}_df_{\hat{a}\hat{b}}^{\;\;\;\dot{c}} = 
- f_{\hat{a}\hat{b}}^{\;\;\;\dot{c}}\,,
\sspace 
\bra t_{\hat{a}} ,t_{\hat{b}} \ket_{\gsf_d} =  
\phantom{-} \bra t_{\hat{a}} ,t_{\hat{b}} \ket_{\gsf}\,,
\label{GHduality} 
\ea 
where ${}_df_{ab}^{\;\;c}$ are the structure constants of $\gsf_d$,
the Lie algebra of $G_d$. For example the cosets ${\rm SL}(n)/{\rm SO}(n)$ 
and ${\rm SU}(n)/{\rm SO}(n)$ are dual in this sense. Using 
(\ref{GHRiemann}) and (\ref{Tlstructure}) an immediate consequence is 
\be 
R_{ijkn}(\mf_d) = - R_{ijkn}(\mf) \,,\sspace 
{}_d\zeta_l = (-)^l \zeta_l\,,\;\; l\geq 1\,,
\label{zetaduality}   
\ee
where $\mf_d$ is the semi-riemannian metric on $G_d/H_d$
and ${}_d\zeta_l$ are the associated coefficients in (\ref{T_gsym}).   
From (\ref{zetaduality}) the table of dual pairs in \cite{Helgason} 
and Hikami's results for compact cosets \cite{Hikami81} one 
obtains table 1 below.

%%%%%%%%%%%%%%%%%%%%%%%%%%%
\vspace{7mm}

%\hspace{-1mm}
\begin{tabular}{|c|c|c|c|}
\hline
$G/H$ & $k$ & $2 \zeta_2$ & $-3 \zeta_3$  \\[2mm] 
\hline 
SL($n\!+\!2$)/SO($n\!+\!2$) & $n$ & $\frac{1}{8}(n\!+\!2)(n\!+\!4)$ & 
$\frac{1}{64}(n+2)(3 n^2 + 22 n + 40)$ \\[4mm] 
$\dfrac{{\rm SU}(n\!+\!1,m\!+\!1)}%
{{\rm S[U}(n\!+\!1)\!\times\! {\rm U}(m\!+\!1)]}$ &  $n\!+\!m$ &
$\frac{1}{2} ( k \!+\! nm \!+\!2)$ & $\frac{1}{16}(k+2) 
(3 k + 3 nm +10)$ \\[4mm]  
$\dfrac{{\rm SO}(n\!+\!2,m\!+\!2)}%
{{\rm SO}(n\!+\!2)\!\times \!{\rm SO}(m\!+\!2)}$ &  $n\!+\!m$ &
$\frac{1}{4} (3 k\! + \!2nm \!+\!4)$ & $\frac{1}{32}[7(k\!+\!4)(k\!+\!2) 
\!+ \!2 nm (3 k\!+ \!14)\! - \!16]$ \\[4mm]  
${\rm SO^*}(2 n\! +\! 4)$/U($n\!+\!2$) & $2n$ & $\frac{1}{2}(n^2 + n +2)$ &
$\frac{1}{8}(3 n^3 + 4 n^2 + 15 n + 10)$\\[4mm]
Sp($2 n$+2,$\R$)/U($n\!+\!1$) & $n$ & $\frac{1}{8} (n^2 + 5 n + 8)$ & 
$\frac{1}{64}( 3 n^3 + 23 n^2 + 72 n + 80)$  \\[2mm]
\hline 
\end{tabular}
%%%%%%%%%%%%%%%%%%%%%%%%%%%%%%%%%%%
\bigskip

Table 1: {\small $l \leq 3$ loop beta function coefficients for 
all non-exceptional cosets arising from (\ref{4Daction}). Here 
$k>0$ except for ${\rm SL}(2)/{\rm SO}(2)$ and $\zeta_1 = - (k+2)/2$.} 
\vspace{4mm}

In the table the riemannian version of the cosets ($\eps_1 = +1$ in 
Eq.~(\ref{mmetric})) was used throughout. The coefficients for the 
pseudo-riemannian version $(\eps_1 =-1$) are identical
\be
\zeta_l(\eps_1 =+1) = \zeta_l(\eps_1 =-1) \,,\sspace l \geq 1\,.
\label{zetalpm}
\ee
For the proof we interpret both symmetric spaces, the $\eps_1 = + 1$
and the $\eps_1 =-1$ version, as different real sections of a complex manifold 
(which may no longer be a symmetric space) and show that the $\zeta_l$ 
are constant for the complex manifold. To this end we promote the 
initially real and positive parameter $q$ in the metric (\ref{mmetric}) 
to a complex one, $q = |q| e^{i\alpha}$, $|q| \neq 0$, $0\leq 
\alpha < 2\pi$. Simultaneously we allow the 
vector fields $A$ to be complex. The complexified metric 
might no longer be that of a symmetric space because the symmetry 
enhancement described after Eq.~(\ref{mmetric}) might cease to 
work. However the generic $\dim \Gbar + 2 k + 2$ isometries of the 
metric (\ref{mmetric}) will still be present, just that the $2k$ 
constant translations $A \ra A + a$ now refer to complex constants 
$a$. These Killing vectors are enough to infer from part (i) of the 
Lemma in appendix B that the curvature scalars $n \zeta_l = 
\mf^{ij} T_{ij}^{(1,l)}(\mf)$, $n = \dim G/H$, in (\ref{T_gsym}) 
are field independent also for the complexified manifold. 
In principle they could depend on the parameter $q$ but since 
the field redefinition $A \ra |q|^{-1/2} e^{-i\alpha/2} A$ removes any 
$q$-dependence in the line element the $\zeta_l$ are $q$-independent. 
On the other hand $A \ra -A$ is an (involutive) isometry and the 
combined sign flip $q \ra -q, A \ra -A$ is equivalent to flipping the 
sign of $\eps_1$ in (\ref{mmetric}). Since the former flip leaves the 
$\zeta_l$ unaffected the latter must do so too, which establishes 
Eq.~(\ref{zetalpm}).

For the exceptional cosets the coefficient $\zeta_1$ directly
follows from the general formula (\ref{zeta1}) quoted earlier: 
\be
\zeta_1 = -2, - \frac{9}{2}, -6\,,-9,\,-15\quad \mbox{for} \quad 
G_2,\,F_4,\,E_6,\,E_7,\,E_8\,,
\ee 
respectively. For each of $G= E_6,\,E_7,\,E_8$, several symmetric 
spaces arise \cite{BM00,CJLP} differing by the non-compact version 
of $G$ and the subgroup $H$ used. Notably $\zeta_1$ is the same 
for the different versions. We have not tried to compute the higher 
order coefficients and are not aware of results in the literature on 
them. 

Let us also briefly mention the special role of hermitian 
symmetric spaces, i.e.~those which admit a complex structure. 
In table 1 these are the $n=0$ member 
of the first series, the $m=0$ members of the  third series, 
and all others \cite{Helgason}. In the present context the 
characteristic feature of hermitian symmetric spaces is that 
they have their `oxidation endpoint' in four dimensions \cite{CJLP}, 
i.e.~the highest possible dimension for a gravity theory which upon 
dimensional reduction gives a 2D theory of the form (\ref{act1}) 
with such a coset $G/H$ (and $h(\rho) =\rho$) is four. The complex 
structure also allows for the introduction of Ashtekar-type variables 
\cite{BrodZag}.

We proceed by illustrating the above features for 
the lowest members of the first three series in table 1, 
which arguably are also the physically most interesting ones. 
The main datum characterizing its 2-Killing vector reduction is 
the metric $\mf$ on the symmetric space $G/H$. This metric is always
of the form (\ref{mmetric}) so that only the constituents of the 
latter have to be specified. Retroactively one can verify that 
a particular metric (\ref{mmetric}) is indeed that of a particular
coset space by examining the Lie algebra generated by its Killing 
vectors; see e.g.~\cite{Galtsov} for the ${\rm SO}(k+2,2)/{\rm SO}(k+2)
\times {\rm SO}(2)$ series. A constructive approach starts by picking 
a unique (e.g.~triangular) representative on the coset $G/H$. After 
choosing an explicit parameterization for it one can use the definition 
of $M$ and (\ref{GHmetric}) to compute the metric. The result
will be of the form (\ref{mmetric}) but with the isometry group 
already identified; see \cite{BM00} for more details on this.   
%%%%%%%%%%%%%%%%%%%%%%%%%%%%%

\underline{ $5$ dimensional gravity:} The cosets are 
${\rm SL}(3)/{\rm SO}(3)$ for $\eps_1 =+1$ and 
${\rm SL}(3)/{\rm SO}(1,2)$ for $\eps_1 =-1$. The Kaluza-Klein vector 
gives rise to two potentials $B,C$; the scalar is denoted by $\vpbar$. 
The scalar-vector coupling is $\mu = \vpbar^2$ and $\nu=0$ with $q=1$; 
the induced metric for $\vpbar$ is $\mfb = \frac{4}{3} \vpbar^{-2}$.
Further $\Upsilon = i\sigma_2$ in terms of the Pauli matrix $\sigma_2$. 
In coordinates $(\Delta,\psi,\vpbar,B,C)$ the metric is 
\be
\mf_{ij} = \frac{1}{\Delta^2} 
\left( \begin{array}{ccccc}
    1    &      0        &   0             &    0            &   0 \\ 
    0    &      1        &   0    & \frac{1}{2}C &  -\frac{1}{2}B  \\ 
    0    &      0        &   \frac{4}{3}\dfrac{\Delta^2}{\vpbar^2} 
& 0 & 0\\    
    0    & \frac{1}{2}C  &   0 & \frac{1}{4}C^2 + \eps_1\Delta \vpbar^2    
& - \frac{1}{4}B C \\
    0    & -\frac{1}{2}B & 0 & - \frac{1}{4}B C 
& \frac{1}{4}B^2 +\eps_1 \dfrac{\Delta}{\vpbar^2} \\ 
\end{array} \right)\,.
\label{SL3metric}
\ee
The first three counter term coefficients in (\ref{Tcounter}) can 
now be computed directly from (\ref{T123}). They come out as 
$\zeta_1 = -3/2,\,\zeta_2 = 15/16, \,\zeta_3 = -65/64$, for 
both signs $\eps_1 =\pm 1$, in agreement with table 1 and 
Eq.~(\ref{zetalpm}).

%%%%%%%%%%%%%%%%%%%%%%%

\underline{Einstein-Maxwell theory:} The relevant cosets are  
${\rm SU}(2,1)/S[{\rm U}(2)\times {\rm U}(1)]$ for $\eps_1 =+1$ and 
${\rm SU}(2,1)/S[{\rm U}(1,1)\times {\rm U}(1)]$ for $\eps_1 =-1$. 
The scalars $\vpbar$ are absent and the Maxwell field is parameterized 
by two potentials $B,C$. The scalar-vector coupling is trivial, 
$\mu = \1_2$ and $\nu = 0$ with $q=4$; further $\Upsilon = i \sigma_2$ 
as above. In coordinates $(\Delta,\psi,B,C)$ the metric reads
\be
\mf_{ij} = \frac{1}{\Delta^2}
\left( \begin{array}{cccc}
1   &      0         &         0        &        0      \\ 
0   &      1         &        2 C       &       -2 B    \\
0   &\phantom{-}2C   &4(C^2+\eps_1\Delta) & - 4 B C       \\
0   &  - 2 B         &   -4 B C         & 4(B^2 +\eps_1\Delta)
\end{array} \right)\,.
\label{SU21metric}
\ee
In contrast to the full theory \cite{DesNieuw74} the 2-Killing vector 
reduction remains strictly renormalizable at one loop. Beyond one 
loop both the full and the reduced theory are non-renormalizable. 
Inserting (\ref{SU21metric}) into (\ref{T123}) gives 
$\zeta_1 = -3/2,\,\zeta_2 = 3/4,\,\zeta_3 = -13/16$, for both signs 
$\eps_1 = \pm 1$.

%%%%%%%%%%%%%%%%%%%%%

\underline{Einstein-Maxwell-dilaton-axion theory:} The 
cosets are ${\rm SO}(3,2)/{\rm SO}(3) \times {\rm SO}(2)$ for 
$\eps_1 = +1$ and ${\rm SO}(3,2)/{\rm SO}(2,1) \times {\rm SO}(2)$ 
for $\eps_1 =-1$. This is the first member of the third series 
in table 1 because the $n=m=0$ system describing the 
2-Killing vector reduction of dilaton-axion gravity corresponds to the 
decomposable coset ${\rm SO}(2,2)/{\rm SO}(2) \times {\rm SO}(2)$. 
It will be treated 
in section 5. In addition to being hermitian symmetric spaces 
all ${\rm SO}(k+2,2)/{\rm SO}(k+2)\times {\rm SO}(2)$, $k \geq 0$, 
have a K\"{a}hler structure detailed in \cite{Galtsov}. 

In the $k=n=1$ system exemplified here the scalar vector couplings
$\mu$ and $\nu$ in (\ref{4Daction}) can be identifies with the 
dilaton $\varphi$ and the pseudo-scalar axion field 
$\chi$, respectively. They also parameterize
the scalar coset $\Gbar/\Hbar = {\rm SL}(2)/{\rm SO}(2)$ with metric 
$\overline{\mf}_{ij} = {\rm diag}(\vp^{-2}, \vp^{-2})$ in coordinates 
$\vpbar^1 = \varphi,\,\vpbar^2 = \chi$. In addition
there are gauge potentials $(B,C)$. With $\Upsilon = i\sigma_2$ 
as in the previous examples and $q=1$ the metric in (\ref{mmetric}) 
is completely determined. Explicitly $\mf_{ij}$ is a $6\times 6$ symmetric 
matrix whose lower $2\times 2$ block reads
\be
\left(
\begin{array}{cc}
\dfrac{\eps_1}{\Delta}\Big(\vp + \dfrac{\chi^2}{\vp} \Big) 
+ \dfrac{C^2}{4\Delta^2} \;\; & 
\dfrac{\eps_1}{\Delta} \dfrac{\chi}{\vp} - 
\dfrac{BC}{4 \Delta^2} \\[4mm]
\dfrac{\eps_1}{\Delta} \dfrac{\chi}{\vp} - 
\dfrac{BC}{4 \Delta^2} & 
\dfrac{\eps_1}{\Delta \vp} + \dfrac{B^2}{4\Delta^2} 
\end{array}
\right)\,,
\ee
in coordinates $\phi^5 = B, \phi^6 = C$. Proceeding as before 
one finds $\zeta_1 = -3/2,\,\zeta_2 = 7/8,\,\zeta_3 = - 89/96$,
once more independent of $\eps_1$ and in agreement with table 1.

%%%%%%%%%%%%%%%%%%%%%%%%%%%%%%%%%%%%%%%%%%%%%%%%%%%%%%%%%%%%%%%%%%%%%%%%%%%
\newpage
\newsection{UV stable fixed point} 

Recapitulating: one can achieve strict cut-off independence in the 
renormalization of dimensionally reduced gravity theories at the 
expense of introducing infinitely many essential couplings --
thereby identifying the systems as non-renormalizable ones. 
Fortunately all these couplings can be arranged into a single 
scalar function $h(\,\cdot\,)$ of one real variable,
whose flow is governed by the beta functional (\ref{betah}). 
We adopt S.~Weinberg's terminology and call a non-renormalizable 
quantum field theory {\it asymptotically safe} if the flow of 
the essential couplings has an ultraviolet stable fixed point. We now 
show that dimensionally reduced gravity theories are asymptotically 
safe in this sense.

Fixed points of the functional $h$-flow correspond to zeros
of the $\betabf_h$ function. They are conveniently found by converting the 
condition $\betabf_h(h) =0$ into the differential equation
\be
\frac{\lb}{2\pi} \rho \dd_{\rho} h = 
C(\lb) h^2 \;\frac{h}{\lb}
\beta_{\lb}\left(\frac{\lb}{h}\right)\,,
\label{betaode}
\ee 
for some $C(\lb) = \sum_{l \geq 0} C_l (\frac{\lb}{2\pi})^l$ with
constant $C_l$. Here $\beta_{\lb} = \lb^2 \dd_{\lb} B_{\lb}$ is the 
conventional beta function (in the minimal subtraction scheme) 
of the symmetric space sigma-model without coupling to gravity. 
The solution corresponding to a $\lb$-independent $C(\lb) = C_0 
= p/\zeta_1$ is 
\be 
h^{\rm beta}(\rho,\lb) = 
\rho^{p} - \frac{\lb}{2\pi} \frac{2\zeta_2}{\zeta_1} 
- \Big(\frac{\lb}{2\pi}\Big)^2 \frac{3\zeta_3}{2 \zeta_1} \rho^{-p}
+ \ldots\,.
\label{betaminsol}
\ee
The fixed point is non-trivial in the sense that the gravitational
degrees of freedom remain self-interacting at the fixed point {\it and} 
remain coupled to the matter degrees of freedom. It is also unique in 
that all other solutions of (\ref{betaode}) violate the boundary 
conditions. Note that the leading quantum correction 
has the scheme independent coefficient $-2 \zeta_2/\zeta_1$.  

In the previous computation $\rho$ was treated as if it was a real
variable. On the other hand $\rho$ is really the renormalized field 
$\rho = \phi^{n+1}$ in Eq.~(\ref{phiren}). Being an other operator 
valued field one would not expect that nonlinear functions
$f(\rho)$ thereof can be built by pointwise multiplication, but rather 
that the latter have to be declared as normal products $\nl f(\rho)\nr$
by `subtracting' the additional short distance singularities. 
Fortunately this is not necessary here as 
\be
\nl f(\rho) \nr = \mu^{d-2} f(\rho)\,,
\label{rhocomposites}
\ee
holds for an arbitrary function $f(\,\cdot\,)$. That is, up to the 
trivial $\mu$-prefactor the function $x \ra f(\rho(x))$ and the 
composite operator can be identified. (Of course this is not true 
for any other field $\phi^j,j \neq n+1$.) Equivalently pointwise 
multiplication of the operator valued field $\rho(x)$ is legitimate. 
The derivation of (\ref{rhocomposites}) parallels that in \cite{PTpaper} 
where also the precise definition of the normal product can be found. 
The upshot here is that $\rho$ can be treated as if it was a classical  
field, rendering both the above and the subsequent computations well-defined.

To proceed we consider the linearization of the flow equation (\ref{hflow}) 
around the fixed point function $h^{\rm beta}$. Since the lowest order term 
is fixed by strict renormalizability an appropriate parameterization is  
\be
\hbbar(\rho,\lb,\mu) = 
h^{\rm beta}(\rho,\lb) + \overline{\delta h}(\rho,\lb,\mu)\,,
\quad  \overline{\delta h}(\rho,\lb,\mu) = 
\frac{\lb}{2\pi} \sbar_1(\varrho,t)
+ \Big(\frac{\lb}{2\pi}\Big)^2 \sbar_2(\varrho,t) + \ldots. 
\label{sbaransatz}
\ee
where the $\sbar_l(\varrho,t)$ are functions of $\varrho:= \rho^p$ 
and $t = \frac{1}{2\pi} \ln \mu/\mu_0$, which vanish for 
$\varrho \ra \infty$ uniformly in $t$. This boundary condition
adheres to the `freezing' of the full nonlinear $\hbbar$-flow
at $\rho = \infty$. For the $t$-dependent quantities the scaling 
dimensions no longer match the powers in $\lb$. A natural grading 
can be introduced as follows: We assign to the renormalization time 
$t$ scaling dimension $1$, and to the functions $\sbar_l$ a scaling 
dimension $1\!-\!l$ in order to match the scaling dimensions of the 
$h_l^{\rm beta}$ under constant rescalings of $h_0 = \varrho$. 
Since the $\sbar_l$ also appear under integrals the appropriate
scaling transformation is 
\ba
(\varrho,\,t) &\rra& (\Lambda \varrho, \,\Lambda t)\,,
\nonum
\sbar_l(\,\cdot\,,\;\cdot\,) & \rra & \Lambda^{1-l}\, 
\sbar_l(\Lambda^{-1} \,\cdot\,,\Lambda^{-1}\,\cdot\,)\,.
\label{scaling}
\ea 
We decompose the linearized flow equations into pieces transforming
with weight $\Lambda^{-l},\,l \geq 1$, under (\ref{scaling}). 
Retroactively one can also restore powers of $\lb$ according to 
this grading. The result is 
\ba
\frac{d}{d t} \dhbar \is 2\pi \int \!\! du \,
\dhbar(u)\,\frac{\delta \betabf_h}{\delta h}
\big(h^{\rm beta}/\lb\big)(u) 
%\nonum
%\is \rho \dd_{\rho} \left[
%- \frac{\zeta_1}{p} \frac{\dhbar}{h} + h \sum_{l \geq 1} l^2 
%\zeta_l \Big(\frac{\lb}{2\pi} \Big)^{l-1} \int_{\rho}^{\infty} 
%\frac{du}{u} \frac{\dhbar(u)}{h(u)^{l+1}} \right]
\nonum 
\is - \frac{\zeta_1}{p h} \rho \dd_{\rho} \dhbar - 
\frac{\dhbar}{h}\sum_{l \geq 2} l(l\!-\!1) \zeta_l 
\Big(\frac{\lb}{2\pi h} \Big)^{l-1} 
\nonum
&+& \rho \dd_{\rho} h \sum_{l \geq 1} l^2 \zeta_l 
\Big(\frac{\lb}{2\pi} \Big)^{l-1} \!\int_{\rho}^{\infty} \frac{du}{u} 
\frac{\dhbar(u)}{h(u)^{l+1}}\,,   
\label{linflow}
\ea 
where in the explicit expression $h$ refers to $h^{\rm beta}$
and we omitted the $\lb$ and/or $\mu$ arguments. 
Decomposing (\ref{linflow}) according to the above grading yields 
a recursive system of inhomogeneous integro-differential equations 
for the $\sbar_l$, $l \geq 1$, 
\be
\frac{d}{d t} \sbar_l =
\zeta_1 \varrho \int_{\varrho}^{\infty} \frac{du}{u^3} \sbar_l(u,t)
- \zeta_1 \dd_{\varrho} \sbar_l
+ R_l[\sbar_{l-1},\ldots,\sbar_1]\,,\quad l \geq 1\,.
\label{slflow}
\ee
Notably the homogeneous parts always have the same form, the 
inhomogeneities however get more complicated with increasing $l$. 
They are differentiable functions of $\varrho$ and $t$ for which we 
write $R_l = R_l(\varrho,t)$. Moreover they come out to have the 
following additive structure: $R_1 =0$ and  
\ba
&& R_l[\sbar_{l-1}, \ldots, \sbar_1] = \sum_{k=2}^l R_{k1}[\sbar_{l+1-k}]
\quad \mbox{for} \quad l \geq 2\quad \mbox{with} \quad
\nonum
&& R_{k1}[\sbar\,] = \rho^{-k}(\alpha_k \sbar + \beta_k \varrho \dd_{\varrho} 
\sbar) + \varrho \int_{\varrho}^{\infty} 
\frac{du}{u^{k+2}} \Big( \sum_{q=0}^{k-1} (u/\varrho)^{q} \gamma_{k,q}\Big) 
\sbar(u,t) \;,
\label{Rk1}
\ea    
where $\alpha_k, \beta_k, \gamma_{k,0}, \ldots,\gamma_{k,k-1}$ are 
real constants, with $\gamma_{k,1} =0$. For example 
\ba
\label{Rk1coeffs}
&\nspace& \\[-7mm]
&\nspace&
\begin{array}{llll}
\alpha_2 = -2\zeta_2\,,\quad & \beta_2 = -2\zeta_2\,,\quad 
& \gamma_{2,0} = 8 \zeta_2 \,,\quad & \mbox{} 
\\[1mm]
\alpha_3 = -\frac{1}{\zeta_1}(8 \zeta_2^2 + 6 \zeta_1 \zeta_3) \,,\;\;
& \beta_3 = -\frac{1}{2\zeta_1}(8 \zeta_2^2 + 3 \zeta_1 \zeta_3) \,,\;\;
& \gamma_{3,0} = 12 \zeta_3 + 36 \frac{\zeta_2^2}{\zeta_1} \,,\;\;
& \gamma_{3,2} = \frac{3}{2} \zeta_3\,.
\end{array}
\nonumber
\ea

Using (\ref{Rk1}) one can recursively construct a solution of 
(\ref{slflow}) satisfying the desired boundary condition. The 
instrumental formula expresses $\sbar_l$ in terms of 
$\sbar_1, \ldots, \sbar_{l-1}$ as 
\begin{subeqnarray} 
\label{slrecursion}
\sbar_1(\varrho,t) \is \varrho \int_{\varrho}^{\infty} 
\frac{du}{u} r_1(u -\zeta_1 t)\;, 
\\
\sbar_l(\varrho,t) \is \varrho \int_{\varrho}^{\infty} \frac{du}{u} 
r_l(u - \zeta_1 t) + \sum_{k=2}^l \int_0^t ds 
F_k[\sbar_{l+1-k}](\varrho,t;s)\;, \quad l \geq 2\,,
\\[4mm]
F_k[\sbar](\varrho,t;s) \is \varrho \int_{\varrho}^{\infty} du 
\frac{\sbar(u - \zeta_1 s, t-s)}{u^2(u - \zeta_1 s)^{k+1}} 
\Big( \gamma_k u + s \zeta_1 (\alpha_k \!+ \!(k\!+\!2)\beta_k) 
- 2 \beta_k (s \zeta_1)^2 u^{-1} \Big)
\nonumber \\
& +&  \frac{1}{(\varrho - \zeta_1 s)^k} 
\Big( \beta_k \,(\varrho -\zeta_1 s) \dd_{\varrho} + 
\alpha_k\!- \!\beta_k \zeta_1 s/\varrho 
\Big)\, \sbar(\varrho - \zeta_1 s, t-s)\,.
\end{subeqnarray} 
Here $r_1$ and $r_l$ are smooth functions of one variable satisfying
$u\, r_k(u) \ra 0$ for $u \ra \infty$; the constants   
$\alpha_k, \beta_k$ and $\gamma_k := \sum_{q=0}^{k-1} \gamma_{k,q}$
refer to the coefficients in (\ref{Rk1}). Since $\sbar_1$ is known
explicitly Eq.~(\ref{slrecursion}) in principle allows one to compute all 
$\sbar_l$, $l \geq 2$, recursively. At each iteration step $l-1 \ra l$ 
a new function $r_l$ enters via the solution of the homogeneous 
Eq.~(\ref{slflow}). Alternatively they can be viewed as parameterizing 
the initial configuration via $r_l(\varrho) = - \varrho \dd_{\varrho} 
[\sbar_l(\varrho, t=0)/\varrho]$. Eventually thus 
$\sbar_l$ is parameterized by $l$ functions of one variable
$r_1,\ldots, r_l$. For example
\begin{subeqnarray}
F_2[\sbar_1] \is -2 \zeta_2 \varrho \int_{\varrho}^{\infty} \frac{du}{u^2} 
\frac{2 u - \zeta_1 s}{(u - \zeta_1 s)^2} r_1(u - \zeta_1 t) 
+ \frac{2 \zeta_2}{\varrho - \zeta_1 s} r_1(\varrho - \zeta_1 t)\,,
\\ 
\sbar_2(\varrho,t) \is \varrho \int_{\varrho}^{\infty} 
\frac{du}{u} r_2(u -\zeta_1 t)
+ \frac{2 \zeta_2}{\zeta_1} \varrho \int_{\varrho}^{\infty} \frac{du}{u} 
\ln(1 - \zeta_1 t/u)\,\dd_u r_1(u -\zeta_1 t)\,.
\label{s12sol}
\end{subeqnarray} 
Based on the recursive solution formula (\ref{slrecursion}) one can 
establish the following result.
\bigskip

{\bf Theorem (UV stability):} Given a smooth initial configuration 
${\mathsf s}_l(\varrho) = \sbar_l(\varrho,0)$, $l \geq 1$,  let 
$\sbar_l(\varrho,t)$ be a solution of (\ref{slflow}) such that for 
$\varrho \ra \infty$ both
$\sbar_l(\varrho,t)$ and $\varrho \dd_{\varrho} \sbar_l(\varrho,t)$ 
vanish uniformly in $t$. Then, for all $l \geq 1$, the 
solution is unique, smooth, and satisfies
\be
\sbar_l(\varrho,t) \rra 0 \quad \mbox{for}\quad t \ra \infty
\quad \mbox{if} \quad \zeta_1 <0 \;, 
\label{UVstability}
\ee
where the convergence is uniform in $\varrho$, for all $\varrho$ bounded 
away from zero. The situation is illustrated in Fig.~1 below. 
\medskip

We do not require that the initial data remain bounded as $\varrho \ra 0$.
For example one might wish to take $\hbbar(\varrho, t=0) = 
\varrho - \frac{\lb}{2\pi} \frac{2 \zeta_2}{\zeta_1}$ as initial 
condition for $\hbbar$. Then $\sbar_1 \equiv 0$ and $\sbar_l(\varrho,t=0) 
\sim \varrho^{-l+1},\;r_l(u) \sim u^{-l}$, $l \geq 2$. 
\bigskip
\bigskip

%%%%%%%%%%%%%%%%%%%%%%%%%%%%%%%%%%%
 % figure sldecay
\begin{figure}[htb]
%\vskip -10mm
\leavevmode
\hskip 45mm
\epsfxsize=70mm
\epsfysize=60mm
\epsfbox{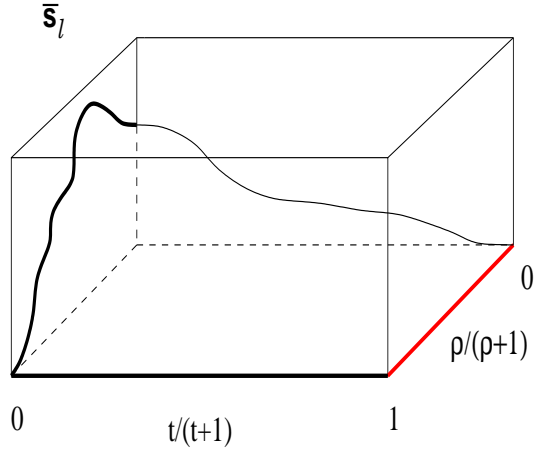}
\caption{{\small Decay of linearized perturbations $\sbar_l(\varrho,t)$ 
for $\zeta_1 <0$: The initial configuration at $t=0$ and the vanishing 
at $\varrho =\infty$ are prescribed. The vanishing for $t \ra \infty$ 
is a dynamical property shown in appendix C.}} 
\label{sldecay}
\end{figure}
%%%%%%%%%%%%%%%%%%%%%%%%%%%%%%%%%%%

If $\zeta_1 >0$ the same result formally holds for $t \ra - \infty$.
However since we are concerned here with UV renormalizability this
`infrared stability' of the UV renormalization flow is 
of little significance. The proof of the theorem is somewhat technical 
and is relegated to appendix C. 

Since $\zeta_1$ enters through the relation $R_{ij}(\mf) = 
\zeta_1 \mf_{ij}$, its value depends on the normalization of $\mf$. 
The relation itself is valid with $\zeta_1 \neq 0$ for 
indecomposable symmetric spaces $G/H$ with $G$ semi-simple
(c.f.~appendix A), but for the higher order analogues in 
(\ref{Tcounter}) $G$ has to be simple in general (c.f.~appendix B).
If $\mf$ has euclidean signature only positive 
rescalings of $\mf$ are allowed and the sign of $\zeta_1$ is intrinsic. 
Then $\zeta_1 <0$ is a necessary and sufficient condition 
for the symmetric space to be non-compact \cite{Helgason}.    
If the metric $\mf$ is indefinite the sign of $\zeta_1$ is 
unambiguously defined only once overall sign flips of $\mf$ are 
prohibited. For the non-compact symmetric spaces arising by 
dimensional reduction this turns out to be the case, moreover 
\be
\zeta_1 = - \frac{k+2}{2} \;,\sspace k = \# \mbox{vector fields} =
\frac{1}{4}(\dim G - \dim \Gbar -3)\,.
\label{zeta1}
\ee
Importantly this is the same for the riemannian ($\eps_1 = +1$) and the 
semi-riemannian ($\eps_1 = -1$) version of the metric in (\ref{mmetric}). 
This means the UV stability of the fixed point holds irrespective 
of the signature of the Killing vectors used in the reduction.   

For the derivation of (\ref{zeta1}) it is enough to note that both 
the overall sign of $\mf$ and the overall normalization 
of the invariant bilinear form $\bra\cdot\,,\,\cdot\ket_{\gsf}$ 
are unambiguously fixed in the symmetric spaces arising by 
dimensional reduction. The result then follows from Eq.~(\ref{GHRiemann}). 
The overall sign of $\mf$ in (\ref{mmetric}) is fixed by the 
requirement of positive energy. In terms of the Lagrangian 
(\ref{4Daction}) this entails that the Kaluza-Klein scalars 
$\vpbar^i$ interact via the positive definite metric 
$\overline{\mf}_{ij}(\vpbar)$ and it also constrains the coupling 
matrix $\mu$ to be positive definite. If there are no Kaluza-Klein
scalars $\vpbar^i$ the two purely gravitational degrees of 
freedom $\Delta,\psi$ serve to fix the overall sign of $\mf$. 
The corresponding $2 \times 2$ block of the metric (\ref{mmetric}) 
always has eigenvalues $(+,+)$, irrespective of the signatures
$\eps_1,\,\eps_2$ of the two Killing vectors, provided the 3D duality 
transformation (\ref{3Dduality}) is performed. For a similar reason 
also the overall normalization of the invariant form 
$\bra\cdot\,,\,\cdot\ket_{\gsf}$ is fixed. This is because in all
cases $G$ is simple and contains the purely gravitational 
$SL(2)$ as a subgroup. Thus up to an overall constant 
$\bra\cdot\,,\,\cdot\ket_{\gsf}$ must coincide with the Killing 
form. Requiring that the restriction of $\bra\cdot\,,\,\cdot\ket_{\gsf}$
to the $sl_2$ subalgebra coincides with the Killing form on 
$sl_2$ fixes $\bra t_a ,\,t_b \ket_{\gsf} = \frac{1}{k+2} f_{ac}^{\;\;\;d} 
f_{bd}^{\;\;\;c}$ \cite{BM00}. In the notation of 
appendix A  this gives $(k+2) \bra t_{\hat{a}} , t_{\hat{b}} \ket_{\gsf} = 2 
f_{\dot{c} \hat{a}}^{\;\;\;\hat{c}}\, f^{\dot{c}}_{\;\hat{c}\hat{b}} = 
-2 R^{\hat{c}}_{\;\,\hat{a}\hat{c} \hat{b}}$, which establishes 
(\ref{zeta1}).

%%%%%%%%%%%%%%%%%%%%%%%%%%%%%%%%%%%%%%%%%%%%%%%%%%%%%%%%%%%%%%%%%%%%%%%%%%%
\newpage 
\newsection{Augmented scalar matter}

The scalar-vector couplings in (\ref{4Daction}) typically arise from 
dimensional reduction of $D \geq 4$ dimensional (super-)gravity 
theories and are largely dictated by the requirement of on-shell covariance. 
Of course one can also couple 4D scalars transforming under some group $\Gt$ 
minimally to 4D gravity without regard to any vectors. 
One obvious case not covered by the systems (\ref{4Daction}) is
when a set of 4D massless free scalars is minimally coupled to 
4D gravity. Another example is 4D dilaton-axion gravity whose 2-Killing 
vector reduction gives rise to the decomposable 
symmetric space ${\rm SO}(2,2)/{\rm SO}(2) \times {\rm SO}(2)$
\cite{IBakas94}. Further, as explained in section 2 the `symmetry enhancement'
from the $\Gbar/\Hbar$ to the $G/H$ coset in the target space 
metric (\ref{mmetric}) hinges on the presence of vector fields 
and certain properties of $\Gbar/\Hbar$. If these conditions are 
not satisfied one can just as well switch off the vectors and consider 
4D gravity minimally coupled to a 4D sigma-model with a coset $\Gt/\Ht$, 
which need no longer be non-compact. For positivity reasons $\Gt/\Ht$ 
should have riemannian signature and we may take $\Gt$ to be simple 
or abelian. The 2-Killing vector reduction of 
these 4D theories gives rise to 2D systems of the form (\ref{act1}) 
where the coset $G/H$ is replaced by the direct product of 
${\rm SL}(2)/{\rm SO}(2)$ with $\Gt/\Ht$.

\newsubsection{Renormalization: coupling to free vs interacting scalars} 

Since it comes at little extra price we consider more generally 
2D sigma-models with target space metric
\be
\Gf_{ij} = \left( \begin{array}{c|c}
(\gf_{ij})_{1\leq i,j\leq n+2} & \mbox{\LARGE 0}\\[2mm]
\hline
\\[-5mm]  
\mbox{\LARGE 0}   & k(\rho) (\mft_{ij})_{1\leq i,j\leq \nbar} 
\end{array}
\right)\,,
\label{Tredmetric}
\ee
where $\gf_{ij}$ is that of Eq.~(\ref{Tmetric}) and $\mft_{ij}$ 
is the metric of some riemannian symmetric space $\Gt/\Ht$ 
of dimension $\nt$ with $\Gt$ simple or abelian (but not necessarily 
non-compact). The total dimension of the target space thus is 
$n+ 2 + \nt$. The function $k(\rho)$ is assumed to be of the form 
\be
k(\rho,\lb) = k_0(\rho) + \frac{\lb}{2\pi} k_1(\rho) + 
\Big(\frac{\lb}{2\pi} \Big)^2 k_2(\rho) + \ldots \,,\sspace
k_0(\rho) \sim \rho^p\,, 
\label{klb} 
\ee
where the condition on $k_0(\rho)$ (with the same $p \neq 0$ as in 
(\ref{hlb})) is motivated by the classical limit and the dimensional 
reduction procedure. The scalar curvature of $\Gf$ is
\be
R(\Gf) = \frac{R(\mf)}{h(\rho)} + \frac{R(\mft)}{k(\rho)}\,,
\label{RicciSG}
\ee
where $R(\mf) = \zeta_1 n$ and $R(\mft) = \zetat_1 \nt$ 
are the constant curvatures of $G/H$ and $\Gt/\Ht$,
respectively. 

Proceeding similarly as in appendix B one finds that the
counter terms have to all loop orders the structure 
\be
T^{(1,l)}_{ij}(\Gf) = \left( \begin{array}{c|c}
(T^{(1,l)}_{ij}(\Gf))_{1\leq i,j\leq n+2} &\mbox{\LARGE 0}\\[2mm]
\hline
\\[-5mm]  
\mbox{\LARGE 0}   & k(\rho)^{1-l} \,\zetat_l\; 
(\mft_{ij})_{1\leq i,j\leq \nt} 
\end{array}
\right)\,,\quad l \geq 1\,,
\label{Tredcounter}
\ee
where in the lower block $l\zetat_l$ are the beta function 
coefficients associated with the $\Gt/\Ht$ sigma-model, 
again defined in minimal subtraction. The upper block has the same form 
as in (\ref{Tcounter}) but the $\rho\rho$ components are now 
differential polynomials in $k(\rho)$ and $h(\rho)$. For example:
\ba
\label{Skl}
T_{\rho\rho}^{(1,1)}(\Gf) \is \frac{n}{\rho^2}S_1(h) + 
\frac{\nt}{\rho^2}S_1(k) + 
\frac{\nt}{2} \dd_{\rho} \ln k \dd_{\rho} \ln h/k\,,
\nonum
T_{\rho\rho}^{(1,2)}(\Gf) \is 0\,,
\\
T_{\rho\rho}^{(1,3)}(\Gf) \is \frac{n}{\rho^2 h^2}S_3(h) + 
\frac{\nt}{\rho^2 k^2} S_3(k) + \frac{\nt\zetat_2}{4 k^2}
\,\dd_{\rho} \ln k \dd_{\rho} \ln h/k\,,
\nonumber
\ea
with $S_l$ as in (\ref{Sl}). We also prepare shorthands for the sums 
$T_{ij}^{(1)}(\Gf) := \!\sum_{l \geq 1} (\frac{1}{2\pi})^l 
T_{ij}^{(1,l)}(\Gf)$, and in particular $\Sf(\rho,\lb) :=
\frac{\rho^2}{h} \sum_{l \geq 1} (\frac{\lb}{2\pi})^l 
T^{(1,l)}_{\rho\rho}(\Gf)$ which generalizes $S(\rho,\lb)$ in 
(\ref{BSdef}).  

For the absorption of these counterterms we 
adopt a similar strategy as before, that is we allow for nonlinear
field renormalizations of the form (\ref{phiren}) a functional 
renormalization (\ref{hren}) of $h$ as well as for its $k$ counterpart, i.e.
\be 
k_B(\rho) = \mu^{d-2} k(\rho,\lb) \left[
1 + \frac{1}{2-d} K(\rho,\lb) + \ldots \right] \,,
\quad K(\rho,\lb) =  \sum_{l \geq 1} \Big(\frac{\lb}{2\pi}\Big)^l 
K_l(\rho)\;.
\label{kren}
\ee
In terms of the diagonal matrix $\Hf = {\rm diag}(H \1_{n+2},
K \1_{\nt}) = \Hf(\rho,\lb)$ the finiteness condition reads
\be
\cL_{\Xi} \Gf_{ij} + (\Hf\,\Gf)_{ij} = 
\lb T^{(1)}_{ij}(\Gf/\lb) \,.
\label{redfiniteness}
\ee
We search for a solution with $\Xi^j = 
(0 ,\ldots, 0, \Xi^{\rho}(\rho,\lb), \Xi^{\sigma}(\rho,\lb), 0,\ldots,0)$, 
where the number of zeros is $n$ and $\nt$, respectively. Again we 
use $\rho = n+1$, $\sigma = n+2$ for the index labeling. The Lie 
derivative term with this $\Xi^j$ is 
\be
\cL_{\Xi} \Gf_{ij} = \left( \begin{array}{c|c}
(\cL_{\Xi} \gf_{ij})_{1\leq i,j\leq n+2} & \mbox{\LARGE 0}\\[2mm]
\hline
\\[-5mm]  
\mbox{\LARGE 0}            & \Xi^{\rho}(h/\lb) \dd_{\rho} k\, 
(\mft_{ij})_{1\leq i,j\leq \nt} 
\end{array}
\right)\,,
\label{LieG}
\ee
where the upper block equals (\ref{Lieg}). The finiteness condition 
(\ref{redfiniteness}) then reproduces the differential equations 
leading to (\ref{finsol}) with $S(\rho,\lb)$ replaced by $\Sf(\rho,\lb)$ and 
one extra equation. Accordingly the solutions for $H$ and $\Xi^{\rho}$ 
in terms of $h$ are the same as in Eq.~(\ref{finsol}) and the solution 
for $\Xi^{\sigma}$ is that in (\ref{finsol}) with $S(\rho,\lb)$ 
replaced by $\Sf(\rho,\lb)$. The extra equation determines -- for 
given $k$ and $h$ -- the counter terms $K$ as 
\be
K\bigg(\frac{h}{\lb}, \frac{k}{\lb}\bigg) = 
\Bt_{\lb}(\lb/k) - \Xi^{\rho} \dd_{\rho} \ln k\,.
\label{Ksol}
\ee
Here  $\Bt_{\lb}(\lb):= \sum_{l \geq 1} (\frac{\lb}{2\pi})^l
\zetat_l$ is the counterpart of $B_{\lb}(\lb)$ in 
Eq.~(\ref{BSdef}) for the $\Gt/\Ht$ sigma-model. 
Further we adjusted the notation for $K$ to stress the 
functional dependence on $h$ and $k$. 

This completes the construction of the basic renormalized Lagrangian 
at some fixed renormalization scale $\mu$. Strict cut-off independence 
can be achieved in terms of the two-fold infinite set of couplings 
$h$ and $k$. The deviation from conventional renormalizability 
in the matter sector is parameterized by $K(h/\lb,k/\lb)$ in 
(\ref{kren}). In contrast to $H$ in (\ref{hren}), (\ref{finsol}) 
$K$ may be non-zero already at one loop. Indeed there are two cases:
\begin{itemize}
\item[(a)] For $\zetat_1 \neq 0$ one has $K_1(\rho) =0$ with $k_0(\rho) = 
\frac{\zetat_1}{\zeta_1} \rho^p$. The matter extended system remains 
strictly renormalizable at one loop.
\item[(b)] For $\zetat_1 =0$ one can achieve 
$k_0(\rho) \sim \rho^p,\,p \neq 0$, only by taking $K_1(\rho) = - \zeta_1 
\rho^{-p}$. Thus already at the 1-loop level standard 
renormalizability cannot be maintained.
\end{itemize}
In case (b) the 4D matter always consists of a 

\underline{Collection of free scalar fields}: 
This is because we assumed $\Gt$ to be either simple or abelian. 
If $\zetat_1=0$ the former possibility is ruled out by the 
results surveyed in appendix A. But if $\Gt$ is abelian not just
the Ricci tensor but the full Riemann tensor vanishes \cite{Loos}; 
in particular then $\zetat_l=0$, $l \geq 1$. The breakdown of 
standard renormalizability in this situation already at one loop seems 
to reflect the well-known feature of full Einstein gravity 
coupled to scalars \cite{tHVelt74} 

Case (a) corresponds to self-interacting scalar matter in 4D
arranged into a $\Gt/\Ht$ sigma-model. The latter can either be put 
in by hand or arise itself via dualization and field redefinitions. 
A typical example for the second possibility is 
  
\underline{Dilaton-axion gravity}: 
The 2-Killing vector reduction is 
described e.g.~in \cite{IBakas94} and leads to a 
${\rm SO}(2,2)/{\rm SO}(2) \times {\rm SO}(2)$ coset.  
Since $so(2,2) \simeq so(1,2) \times so(2,1)$ is semi-simple rather 
than simple this theory cannot directly be subsumed into the class 
of systems in section 2 (though it can be viewed as a limiting case 
e.g.~of the ${\rm SO}(k+2,2)/{\rm SO}(k+2)\times {\rm SO}(2)$ series 
in table 1 with $k=0$ vectors). Denoting the dilaton by $\varphi$ 
and the pseudo-scalar `axion' field by $\chi$ the target space 
metric is of the form (\ref{Tredmetric}) with 
\ba
G/H = {\rm SL}(2)/{\rm SO}(2): &\quad & 
\mf = {\rm diag}(\Delta^{-2},\Delta^{-2})\,,
\nonum
\Gt/\Ht = {\rm SL}(2)/{\rm SO}(2): &\quad & 
\widetilde{\mf} = {\rm diag}(\vp^{-2},\vp^{-2})\,,
\label{O22metric}
\ea
in coordinates $(\Delta,\psi,\rho,\sigma,\varphi,\chi)$. In particular 
$\zeta_l = \zetat_l$, $l \geq 1$, which entails that one can
consistently take $h(\rho) = k(\rho)$; c.f.~below. The counter term
formulas (\ref{Tcounter}) and (\ref{Tredcounter}) then coincide, 
which is a special case of part (iii) of the Lemma in appendix B. 
Explicitly the first three coefficients are $\zeta_1 = -1,\,
\zeta_2 = 1/2,\,\zeta_3 = -5/12$, and coincide with the ones for 
pure gravity. 

The qualitative differences between self-interacting scalars 
(case (a), $\zetat_1 \neq 0$) and free scalars (case (b), 
$\zetat_1 = 0$) become more pronounced if one considers 
the renormalization flow of the generalized coupling $k$.

%%%%%%%%%%%%%%%%%%%%%%%%%%%%%%%%%%%%%%%%%%%%%%%%%%%%%%%%%%%%%%%%%%%%%
\newsubsection{{\boldmath $k$}-flow and its fixed point structure}  

For functionals $X$ of $h$ and $k$ it is convenient to introduce the 
scaling operator 
\be
\dot{X}(h,k) = \int du \left( 
h(u)\frac{\delta X}{\delta h(u)} + k(u)\frac{\delta X}{\delta k(u)}   
\right)\,.
\label{dotdef}
\ee
In particular we prepare 
\be
\dot{\Xi}^{\rho}(h/\lb) = - \rho \int_{\rho}^{\infty} 
\frac{du}{u} \frac{h(u)}{\lb} \beta_{\lb}\left( \frac{\lb}{h(u)}\right)\,.
\label{Xidot}
\ee
Using (\ref{Ksol}) in (\ref{kren}) one obtains the flow equation
\ba
\mu \frac{d}{d \mu} \kbar  \is  \lb \betabf_k
\bigg(\frac{\hbbar}{\lb},\frac{\kbar}{\lb}\bigg)\,,
\nonum
\lb \betabf_k \bigg(\frac{h}{\lb},\frac{k}{\lb}\bigg)
\is (2-d) k - k \dot{K} = 
(2-d) k + \frac{k^2}{\lb} \betat_{\lb}\bigg(\frac{\lb}{k}\bigg) 
+ \dot{\Xi}^{\rho}\bigg(\frac{h}{\lb}\bigg) \dd_{\rho} k \,,
\label{kflow}
\ea
where $\betat_{\lb} = \lb^2 \dd_{\lb} \Bt_{\lb}$ is the beta 
function of the $\Gt/\Ht$ model. Since we insist on $k_0(\rho) \sim \rho^p$ 
a natural boundary condition for Eq.~(\ref{kflow}) is 
$\kbar(\rho,t)/\rho^p \ra \zetat_1/\zeta_1$, as $\rho \ra \infty$.

The $\dot{\Xi}^{\rho}$ term in (\ref{kflow}) describes the impact 
of the gravitational $G/H$ model on the scalar matter in 
$\Gt/\Ht$. For comparison note that using $H = - \Xi^{\rho} 
\dd_{\rho} \ln h + B_{\lb}(\lb/h)$ from (\ref{finsol}) 
the $\betabf_h$ function can be written in a similar form: $-h \dot{H} = 
\frac{h^2}{\lb} \beta_{\lb}(\frac{\lb}{h}) + \dot{\Xi}^{\rho} 
\dd_{\rho} h$. Evidently the key difference is that the 
$\hbbar$-flow equation is autonomous while the $\kbar$-flow is 
triggered by $\hbbar$. In the special case when $\zeta_l = 
\zetat_l,\,l \geq 1$, one sees that $\hbbar = \kbar$ is a 
solution of the coupled system of flow equations. In particular 
this entails that the results in sections 3 and 4 generalize from
cosets $G/H$ with $G$ simple to product manifolds $G/H = 
\Gt/\Ht \times \Gt/\Ht$ containing two identical factors (or factors
with the same coefficients $\zetat_l,\, l \geq 1$) with 
$\Gt$ simple. An example where this is relevant is the before 
mentioned dilaton-axion
gravity which may be subsumed either under the present framework 
with $G/H = \Gt/\Ht = {\rm SL}(2)/{\rm SO}(2)$, or under the 
generalized framework of sections 3,4 with coset $G/H = 
{\rm SO}(2,2)/{\rm SO}(2) \times {\rm SO}(2)$. Both 
descriptions yield the same results.

Returning to the general case one would expect that the relevant stationary 
point of the $\kbar$-flow is the one where also $\hbbar$ is at its fixed 
point. Indeed, as we shall see later, the trace anomaly vanishes modulo 
an improvement term iff $\betabf_h(h) =0 = \betabf_k(h,k)$. At $d=2$ the 
latter gives the condition
\be
\frac{\lb}{2\pi} \rho \dd_{\rho} k = C(\lb) h^{\rm beta} 
\,\frac{k^2}{\lb} \betat_{\lb} \bigg(\frac{\lb}{k}\bigg) \,,
\label{kbetaode}
\ee
where $C(\lb)$ is the same constant as in (\ref{betaode}) and we 
used 
\be
\dot{\Xi}^{\rho}(h^{\rm beta}/\lb) = - \frac{\lb}{2\pi C(\lb)} 
\frac{\rho}{h^{\rm beta}(\rho)}\,.
\label{Xidotbeta}
\ee
As before we take $C(\lb) = p/\zeta_1$ and write $k^{\rm beta}(\rho,\lb) 
= \sum_{l \geq 1} (\frac{\lb}{2\pi})^l k_l^{\rm beta}(\rho)$ for the 
solution. For theories with $\zetat_1 \neq 0$ the first two
terms are
\be
k_0^{\rm beta}(\rho) = \frac{\zetat_1}{\zeta_1} \rho^p \,,
\sspace 
k_1^{\rm beta}(\rho) = 2 \left(\frac{\zetat_2}{\zetat_1} -
\frac{\zeta_2 \zetat_1}{\zeta_1^2}\right) \ln\rho^p 
- \frac{2 \zeta_2}{\zeta_1}\,.
\label{k01beta}
\ee
The additive constant in $k_1^{\rm beta}$ is chosen such that for 
$\zetat_2/\zetat_1^2 = \zeta_2/\zeta_1^2$ it coincides with 
$h_1^{\rm beta}$. Generally the $k_l^{\rm beta}$ contain terms of the 
same form as $h_l^{\rm beta}$ with coefficients which for generic 
$\zetat_l,\,\zeta_l$ are not fully determined. We fix them by 
requiring that in the special case where $\zetat_l =\zeta_l$, 
$l \geq 1$, one has $h^{\rm beta} = k^{\rm beta}$. Inserting 
(\ref{betaminsol}), (\ref{k01beta}) into $K$ one finds  
\be
K(\rho,\lb) = \Big(\frac{\lb}{2\pi}\Big)^2 \left( \frac{\zeta_2}{2} 
- \frac{\zetat_2 \zeta_1^2}{\zetat_1^2} \right) \rho^{-2 p} 
+ O(\lb^3) \,,\quad \mbox{for} \quad h= h^{\rm beta}\,,\;k= k^{\rm beta}\,.
\ee
In general therefore conventional renormalizability is lost 
at two loops even at the fixed point.  

On the other hand for theories with $\zetat_1 =0$ the $\betat_{\lb}$ function 
vanishes identically, one is dealing with free scalars, and 
(\ref{kbetaode}) implies $k= {\rm const}$. 
However this is in conflict with the requirement $k_0(\rho) 
\sim \rho^p,\,p \neq 0$, needed to allow for an interpretation as 
a dimensionally reduced system. In fact for constant $k$ the matter sector     
decouples from the gravitational sector. We conclude that when a 
collection of {\it free} massless fields is minimally coupled to 4D Einstein 
gravity the 2-Killing vector reduction has only a {\it degenerate 
fixed point} where the scalars decouple from gravity.

This illustrates that the non-trivial gravitational fixed point found 
in \cite{PTpaper} --  and its matter generalizations obtained here -- 
are not automatic consequences of the relaxed notion of 
(`conformal') renormalizability. For example if the `backreaction' 
terms proportional to $\dot{\Xi}^{\rho}$ were dropped in the 
$\hbbar$ and $\kbar$ flow equations only the trivial fixed point
$k = h = 0$ would exist, presumably corresponding to the absence 
of interaction. Notably this does not happen for gravity coupled 
to abelian gauge fields and self-interacting scalars. It is only for 
free scalars that the flow equation (\ref{kflow}) becomes linear
and the $\dot{\Xi}^{\rho}$ term precludes the existence 
of a genuine fixed point. Moreover, if one relies on the 
principle \cite{PTpaper} that an UV fixed point in the reduced theory 
should be a prerequisite for an UV fixed point in the full theory, 
the present result suggests a similar `triviality' result for 
4D Einstein gravity minimally coupled to free massless scalars --
that is the gravity-matter coupling is expected to vanish at the fixed 
point. Although preliminary investigations in the full theory based on 
non-perturbative flow equations are available \cite{PercPern02}
the issue so far does not seem to have been studied.

Let us return to the case where the augmented 4D scalars are 
self-interacting and its 2-Killing vector reduction has 
$\zetat_1 \neq 0$ in which case a non-trivial fixed point exists. 
In order to investigate its stability properties we 
linearize (\ref{kflow}) around $(h^{\rm beta}, k^{\rm beta})$.
We parameterize $\overline{\delta h}$ as in (\ref{sbaransatz}) 
and $\overline{\delta k}$ analogously as 
\be
\kbar(\rho,\lb,\mu) = 
k^{\rm beta}(\rho,\lb) + \overline{\delta k}(\rho,\lb,\mu)\,,
\quad  \overline{\delta k}(\rho,\lb,\mu) = 
\frac{\lb}{2\pi} \kappabar_1(\varrho,t)
+ \Big(\frac{\lb}{2\pi}\Big)^2 \kappabar_2(\varrho,t) + \ldots. 
\label{kappabaransatz}
\ee
Here $\varrho = \rho^p$ and $t = \frac{1}{2\pi} \ln \mu/\mu_0$, as before.
The perturbations $\kappabar_l(\varrho,t)$ are supposed to vanish 
for $\varrho \ra \infty$ uniformly in $t$ and are 
assigned scaling dimension $1\!-\!l$ as in (\ref{scaling}). Decomposing the 
linearized $\kbar$-flow according to this grading and restoring powers 
of $\lb$ yields
\ba
\frac{d}{dt} \overline{\delta k} \is 2\pi \int \! du\, 
\frac{\delta \betabf_k(h^{\rm beta}/\lb,k^{\rm beta}/\lb)}{\delta k(u)}
\,\overline{\delta k}(u) 
\nonum
\is - \frac{\zeta_1}{p} \frac{1}{h} \rho \dd_{\rho}( \overline{\delta k})
 - \sum_{l \geq 2} l(l\!-\!1) \zetat_l \Big(\frac{\lb}{2\pi} \Big)^{l-1}
k^{-l} \overline{\delta k} 
\nonum
&& + \rho \dd_{\rho} k \sum_{l \geq 1} l^2 \zeta_l   
 \Big(\frac{\lb}{2\pi} \Big)^{l-1} \!\int_{\rho}^{\infty} \frac{du}{u} 
\frac{\overline{\delta h}(u)}{h(u)^{l+1}}\,,
\label{klinflow}
\ea
where in the explicit expression $h$ and $k$ refer to $h^{\rm beta}$ 
and $k^{\rm beta}$. Comparing with (\ref{linflow}) one sees that 
in the special case when $\zeta_l = \zetat_l,\,l \geq1$, and 
hence $h^{\rm beta} = k^{\rm beta}$, also $\overline{\delta h} = 
\overline{\delta k}$ is a solution of the combined linearized flow 
equations. In the following we exclude this trivial situation
and consider $\overline{\delta h}$ and  $\overline{\delta k}$
as independent perturbations, where $\overline{\delta h}$ is 
supposed to be known through the solution of (\ref{linflow}).   

Then Eq.~(\ref{klinflow}) converts into a recursive system of 
inhomogeneous partial differential equations for the $\kappabar_l$,
$l \geq 1$. The first equation is 
\be
\frac{d}{dt} \kappabar_1 + \zeta_1 \dd_{\varrho} \kappabar_1 
= \zetat_1 \varrho \int_{\varrho}^{\infty} \frac{du}{u^3}\, 
\sbar_1(u,t)\,.
\label{kappa1flow}
\ee
Specifying initial data by some function $\jmath_1$ of one variable 
vanishing at infinity the solution of (\ref{kappa1flow}) is 
\be
\kappabar_1(\varrho,t) = \jmath_1(\varrho - \zeta_1 t) + 
\frac{\zetat_1}{\zeta_1}
[ \sbar_1(\varrho,t) - \sbar_1(\varrho - \zeta_1 t, 0)]\,.
\label{kappa1}
\ee
Evidently the condition $\zeta_1 <0$ ensures that $\kappabar_1(\varrho,t)$
vanishes for $t \ra \infty$ uniformly in $\varrho$. Notably this 
holds irrespective of the sign of $\zetat_1$, only $\zetat_1 \neq 0$ 
is required. The higher order equations are of the form 
\be
\frac{d}{dt} \kappabar_l + \zeta_1 \dd_{\varrho} \kappabar_l 
= \zetat_1 \varrho \int_{\varrho}^{\infty} \frac{du}{u^3} \sbar_l(u,t) + 
Q_l[\kappabar_{l-1},\ldots, \kappabar_1; \sbar_{l-1}, \ldots \sbar_1]\,,
\quad l\geq 2\,,
\label{kappalflow}
\ee
where $Q_l$ has a similar additive structure as $R_l$ in (\ref{Rk1}). 
Regarding the right hand side as an in principle known function
$Q_l(\varrho,t)$ the solution with initial data $\kappabar_l(\varrho, 0) 
= \jmath_l(\varrho)$ is 
\ba
\kappabar_l(\varrho,t) \is \jmath_l(\varrho - \zeta_1 t) +
\frac{\zetat_1}{\zeta_1}
[ \sbar_l(\varrho,t) - \sbar_l(\varrho - \zeta_1 t, 0)] 
\nonum
&+& \int_0^t ds \,\Big(Q_l - \frac{\zetat_1}{\zeta_1}R_l\Big)
(\varrho - \zeta_1 s, t-s) \,.
\label{kappal}
\ea
From here it is relatively straightforward to establish the desired 
uniform decay properties for $\varrho  \ra \infty$ (boundary condition) 
and $t \ra \infty$ (renormalization group dynamics). The analysis is simpler 
than the one in appendic C because the $\kappabar_p$, $p=1,\ldots,l\!-\!1$,
enter without integrals in $Q_l$ and the $\sbar_p$ are already 
known to decay. We thus omit the details and just state that
$\kappabar_l(\varrho, t) \ra 0$ for $t \ra \infty$ 
for $\zeta_1 <0$, uniformly in $\varrho$ and irrespective of the 
sign of $\zetat_1 \neq 0$. The case $\zetat_1 =0$ has already been 
discussed and we can summarize the results:   
\bigskip

{\bf Theorem (free vs interacting 4D scalars)} 
\vspace{-6mm} 

\begin{itemize} 
\item[(a)] For 4D scalar matter corresponding 
to a $\Gt/\Ht$ sigma-model with $\zetat_1 \neq 0$ in the 
reduced theory a non-trivial fixed point $(h^{\rm beta}, k^{\rm beta})$ 
exists where gravity remains self-interacting and coupled to the scalars. 
The fixed point is UV stable under the two-fold infinite set 
of perturbations $(\sbar_l, \kappabar_l)$, $l \geq 1$. An exact analogue 
of the theorem in section 4 holds for both $\sbar_l$ and $\kappabar_l$, 
irrespective of the sign of $\zetat_1 \neq 0$.     
\item[(b)] If $\zetat_1 =0$ the 4D matter consists of a 
collection of free scalar fields. Only a `trivial' fixed 
point exists where the scalars decouple from gravity.
\end{itemize}
%
%%%%%%%%%%%%%%%%%%%%%%%%%%%%%%%%%%%%%%%%%%%%%%%%%%%%%%%%%%%%%%%%%%%%%%%%%
\newpage
\newsection{Conformal invariance co-exists with local running couplings} 

So far we took for granted that the vanishing conditions
$\betabf_h(h) =0$ and $\betabf_k(h,k)=0$ are related to an intrinsic 
property of the system. Indeed, as we verify here, the simultaneous 
vanishing of both beta functionals is a necessary and sufficient condition 
for the trace anomaly to vanish modulo an improvement term. Specifically 
the result is: 
\ba
&& \nl T^{\mu}_{\;\,\mu} \nr = \dd^{\mu} \dd_{\mu}\nl \Phi \nr \quad
\Longleftrightarrow \quad \betabf_h(h) =0=\betabf_k(h,k)\,, 
\nonum
&& \mbox{with} \quad \Phi = \frac{\lb}{2\pi} \frac{1}{C}(a \ln \rho + 
b \sigma) - \int_{\rho}^{\infty} du W_{\!\rho}(\rho) - 
b \int^{\rho} \frac{du}{u} h(u) \dot{\Xi}^{\sigma}(u)\,.
\label{trace1}
\ea 
Here $\nl T_{\mu\nu}\nr$ is the renormalized energy momentum 
tensor based on (\ref{constr}) and $W_i$ is a co-vector induced by 
operator mixing which is built from the curvature tensors of $\Gf_{ij}$
in (\ref{Tredmetric}) and their covariant derivatives. $W_i$ can be shown 
to have the form 
\ba
W_i(\phi) \is (0,\ldots, 0, W_{\!\rho}(\rho), 0, 0, \ldots, 0)\,,
\nonum
W_{\!\rho}(\rho) \is \Big(\frac{\lb}{2\pi} \Big)^3 \frac{1}{8} 
\dd_{\rho} \!\left( \frac{n\zeta_2}{h^2} + 
\frac{\nt\zetat_2}{k^2} \right) + O(\lb^4)\,,
\label{W}
\ea
where the number of zeros is $n$ and $\nt +1$, respectively. 

On the other hand the renormalized energy momentum tensor is 
unique only up to addition of an improvement term $\nl \tau_{\mu\nu} \nr$
with potential $\tau(\phi)$. In order to qualify as an improvement potential 
compatible with the ``conformal renormalizability'' requirement adopted, 
$\tau$ has to be a function of $\rho$ and $\sigma$ only with $\sigma$ 
entering linearly \cite{PTpaper}, viz
\ba
\nl \tau_{\mu\nu}\nr  \is (\dd_{\mu}\dd_{\nu} - \eta_{\mu\nu} \dd^2) 
\nl \tau(\phi)\nr \,,
\nonum
\tau(\phi) \is  f(\rho,\lb) + \tau_0(\lb) \sigma\,.
\label{Phi}
\ea
Clearly its trace is of the same form as (\ref{trace1}) so by defining
$T^{\rm imp}_{\mu\nu} := T_{\mu\nu} + \tau_{\mu\nu}$, with $\tau := \Phi$, 
one can render the improved energy momentum tensor traceless. 
Inserting the expression for $\dot{\Xi}^{\sigma}$ derived from 
(\ref{finsol}) (with $S$ replaced by $\Sf$) this fixes 
\be
\tau_0(\lb) = \frac{\lb}{2\pi} \frac{b}{C(\lb)}\,,\sspace 
\dd_{\rho} f = \frac{\lb}{2\pi} \frac{a}{2 C} \frac{1}{\rho} 
+ W_{\!\rho}(\rho) - \frac{h}{2\rho} \int^{\rho} \frac{du}{u} 
\dot{\Sf}(u,\lb) \,. 
\label{fsol}
\ee  
With the previous choice $C(\lb) = p/\zeta_1$ the improvement potential
(\ref{Phi}) thus is completely determined at the fixed point 
up to some inessential integration constants. Away from the fixed point 
the function $f(\,\cdot\,)$ can be shown to be subject to a
renormalization flow equation which in principle determines it
in terms of the running $\hbbar$ and $\kbar$. We omit the details,
c.f.~\cite{PTpaper}.

It remains to derive Eq.~(\ref{trace1}). To this end one operates
with $1 - h \frac{\delta}{\delta h} - k \frac{\delta}{\delta k}$
on the finiteness condition (\ref{redfiniteness}). Using the scaling 
properties of the constituents 
\be
\dot{T}^{(1,l)}_{ij}(\Gf) = (1-l) T^{(1,l)}_{ij}(\Gf)\,,\quad     
\dot{\phi}_l^j = - l \phi_l^j \,,\quad 
\dot{H}_l = - l H_l\,,\quad 
\dot{K}_l = - l K_l\,,
\label{fscaling} 
\ee
one finds 
\be
\sum_{l \geq 1} \Big(\frac{\lb}{2\pi} \Big)^l l T^{(1,l)}(\Gf) 
= - \dot{\Hf} \Gf - \cL_{\dot{\Xi}} \Gf\,,
\label{dotredfiniteness}  
\ee
with $\dot{\Hf}={\rm diag}(\dot{H}\1_{n+2},\dot{K}\1_{\nt})$.
The left hand side enters the general formula for the trace anomaly 
of a riemannian sigma-model \cite{Tseytlin87,Osb87,CurciPaff}.
Inserting (\ref{dotredfiniteness}) the 
anomaly can be rewritten as 
\be
\nl T^{\mu}_{\;\,\mu} \nr = - \frac{1}{2} \nl (\dot{\Hf} \Gf)_{ij} 
\dd^{\mu} \phi^i \dd_{\mu} \phi^j \nr + 
\dd^{\mu} \nl \dd_{\mu} \phi^i(W_i - \dot{\Xi}_i) \nr\,.
\label{trace2}
\ee
This is of the form (\ref{trace1}) iff $\dot{H} = \dot{K} =0$ and 
$W_i - \dot{\Xi}_i$ is the gradient of a scalar. Combining (\ref{W}) 
with the form of $\dot{\Xi}_i = \Gf_{ij} \dot{\Xi}^j$ and 
(\ref{Xidotbeta}) one finds that at the fixed point indeed 
$W_i - \dot{\Xi}_i = \dd_i \Phi$, with $\Phi$ as in (\ref{trace1}). 

The improved energy momentum tensor is traceless at the fixed point 
so that its components $\nl \cH_0 \nr := \nl T^{\rm imp}_{00} \nr = 
\eps_1 \eps_2 \nl T^{\rm imp}_{11} \nr$ and $\nl \cH_1 \nr := 
\nl T^{\rm imp}_{01} \nr$ can be interpreted as quantum versions of 
the hamiltonian and 1D diffeomorphism constraints, respectively, 
see Eq.~(\ref{constr}). The linear combinations $\nl \cH_0 \pm \cH_1 \nr$ 
are thus expected to generate two commuting copies of a Virasoro algebra
with formal central charge $c = 2 + \dim G/H$. This central charge 
is only formal because it refers to a state space with indefinite 
norm. The construction of quantum observables commuting with 
the constraints and the exploration of the physical state space    
are major desiderata.

There is a conceptual tension between the expectation that a 
quantum theory of gravity should in some sense be diffeomorphism 
invariant and the intuition that in a matter coupled theory the 
conventional running couplings should leave a remnant at the 
gravitational fixed point. An interesting lesson to be learned 
from the dimensionally reduced gravity theories is that both 
intuitions can elegantly be reconciled.   
 
To this end one combines the flow equations for the 
essential couplings $\hbbar$ and $\kbar$ with the flow 
for the inessential $\rho$ `coupling' \cite{PTpaper}
\be
\mu \frac{d}{du} \rhobar = - \dot{\Xi}^{\rho}[\hbbar/\lb](\rhobar)\,, 
\label{rhoflow}
\ee
which generalizes the scale dependence carried by the wave function 
renormalization constant in a multiplicatively renormalizable quantum 
field theory. Evaluating the running coupling functions $\hbbar$ and 
$\kbar$ at the `comoving' field $\rhobar$ yields quantities
\be
\lbbar_h(\mu) := \frac{1}{\hbbar(\rhobar,\mu)} \,,\sspace 
\lbbar_k(\mu) := \frac{1}{\kbar(\rhobar,\mu)} \,,
\label{lbhk} 
\ee
that depend parametrically on the {\it value} of $h(\rho(x))$ and $k(\rho(x))$ 
-- and hence on $x$ -- and which describe the running of these (inverse) 
values. Combining (\ref{hflow}), (\ref{kflow}) with (\ref{rhoflow}) one 
obtains 
\be
\mu \frac{d}{d\mu} \lb \lbbar_h = - \beta_{\lb}(\lb \lbbar_h) \,,
\sspace 
\mu \frac{d}{d\mu} \lb \lbbar_k = - \betat_{\lb}(\lb \lbbar_k) \,.
\label{lbhkflow}
\ee  
These are the usual flow equations for the one coupling $G/H$ and 
$\Gt/\Ht$ sigma-models, respectively! In other words 
the `gravitationally dressed' functional flows for $\hbbar$ and $\kbar$ have 
been `undressed' by reference to the scale dependent `rod field' 
$\rhobar$. The equations (\ref{lbhkflow}) are not by themselves useful 
for renormalization purposes -- which requires determination of 
the flow of $\hbbar(\,\cdot\,,\mu)$ and $\kbar(\,\cdot\,,\mu)$ 
with respect to a {\it fixed} set of field coordinates. Moreover in 
the technical sense $\lbbar_h$ and $\lbbar_k$ are ``inessential''
couplings. However since Eqs.~(\ref{lbhkflow}) are valid for any 
$\hbbar$ and $\kbar$, in particular for their fixed points 
$h^{\rm beta}$ and $k^{\rm beta}$, they display how 
diffeomorphism (i.e.~here 2D conformal) invariance 
can co-exist with conventional scale dependent running parameters.

%%%%%%%%%%%%%%%%%%%%%%%%%%%%%%%%%%%%%%%%%%%%%%%%%%%%%%%%%%%%%%%%%%%%%%%%%

\newsection{Conclusions}

The asymptotic safety property offers for a non-renormalizable quantum 
field theory similar rewards as asymptotic freedom does for a 
renormalizable one. Here a large class of such theories -- arising 
as the 2-Killing vector reduction of 4D Einstein gravity coupled to 
scalars and abelian gauge fields -- has been perturbatively constructed.
The non-renormalizability manifests itself in the presence of infinitely 
many essential couplings $\lbbar_n, \,n\geq 0$, which (morally though not
technically) parameterize the function $\hbbar(\varrho,t) = 
\varrho + \sum_{n \geq 0} \lbbar_n(t) \varrho^{-n}$ featuring in 
section 4 (and similarly for $\kbar$ used in section 5). 
Here $\varrho$ can be identified with the ``area radius'',
at some fixed renormalization scale $t_0$, defined locally by the two 
Killing vector fields. The running couplings  $\lbbar_n$ approach 
a non-trivial fixed point in the ultraviolet, i.e.~$\lbbar_n(t) \ra \lb_n^*$
as $t \ra \infty$. As mentioned in the introduction this is a
surprising feature for which no obvious explanation exists in 
the reduced theories. On the other hand the existence of this fixed 
point can be argued to be a prerequisite for the full theories to 
have an UV stable fixed point. 

A running coupling is not in itself a physical quantity, although 
the nature of the running of course has an impact on the latter. 
In a diffeomorphism invariant theory it is not obvious to which 
quantities one should attribute the status of `physical quantity'. 
Uncontroversial candidates are ``classical observables'', i.e.~quantities 
weakly Poisson commuting with the hamiltonian and the diffeomorphism 
constraints. Remarkably, in the above reduced theories an infinite 
set of such observables can classically be constructed. An important 
open problem therefore is to investigate their fate in the quantum 
theory (e.g.~in the spirit of \cite{Lusch}) and the impact the 
asymptotic safety property potentially has 
on their relations. This might help to characterize the asymptotic 
safety property also in the full theories in terms of physical quantities,
and thus in a way that is independent of a particular computational and 
conceptual approach to quantum gravity. 
\bigskip

{\bf  Acknowledgments:}
I wish to thank M.~Reuter, P.~Forg\'{a}cs and T.~Duncan for discussions. 
This work was completed while visiting the University of Pittsburgh, 
where it was supported in part by NSF grant PHY00-88946. 

%%%%%%%%%%%%%%%%%%%%%%%%%%%%%%%%%%%%%%%%%%%%%%%%%%%%%%%%%%%%%%%%%%%%%%%%%
\newpage 

\setcounter{section}{0}
\newappendix{Symmetric space sigma-models}

Symmetric space sigma-models describe the dynamics of generalized 
harmonic maps from a base manifold $\Sigma$ to an indecomposable symmetric 
space (as defined below) of the form $G/H$, where $G$ is the real form 
of a semi-simple 
Lie group with Lie algebra $\gsf$ and $H$ is a maximal subgroup of 
$G$ with Lie algebra $\hsf$ \cite{ZJbook,BHZJ80}. For the exposition 
in this appendix the nature of the base manifold $\Sigma$ is inessential, 
for definiteness we take $\Sigma = \R^d$, $d \geq 1$, and denote the 
coordinates by $x^{\alpha}$. There are two useful action principles for 
these coset sigma-models. The first is a gauge theoretical one 
\begin{equation}
S[\cV, Q] = \frac{1}{2} \int d^d x\, \bra D^{\alpha} \cV
\cV\inv , \,D_{\alpha} \cV\cV\inv \ket_{\gsf}\;,
\label{sact1}
\end{equation} 
where $\cV$ is a group-valued field transforming as $\cV \ra \cV h$
under an $H$-valued gauge transformation and $Q_{\alpha}$ is the
associated connection ensuring that $D_{\alpha} \cV = \partial_{\alpha} \cV
- \cV Q_{\alpha}$ transforms covariantly. The currents 
$D_{\alpha} \cV \cV^{-1}$ 
take values in the Lie algebra $\gsf$ and $\bra\cdot\,,\,\cdot\ket_{\gsf}$ 
is a non-degenerate bilinear form on it invariant under ${\rm Ad}G$ and 
the (differential of the) involution $\tau$ below.  Clearly the gauge 
symmetry removes $\dim H$ degrees of freedom leaving $\dim G/H$ physical ones.
One can fix a gauge and work with preferred representatives $\cV_{\!*}$ 
intersecting each orbit once. Then $G$ acts nonlinearly via 
$\cV_{\!*} \mapsto g \cV_{\!*}\, h(\cV_{\!*},g)$,
$g \in G$, on the representatives because a gauge transformation is 
needed to bring $g \cV_{\!*}$ back into the original section.

Alternatively one can obtain a non-redundant parameterization of the
coset space by matrices $M \in G$ obeying a suitable quadratic
constraint $M \tau_0(M) = \pm\1$, where $\tau_0$ is an involutive
outer automorphism of $G$.  The subgroup $H$ can then be characterized
as the set of fixed points of a related involutive automorphism $\tau$ 
of $G$, given by $\tau(g) = g_0\inv \tau_0(g) g_0$, for some fixed $g_0 \in G$
which likewise satisfies $g_0 \tau_0(g_0) = \pm\1$. (More precisely 
$(G_{\tau})_0 \subset H \subset G_{\tau}$, where $G_{\tau}$ is the 
fixed point set of $\tau$ and $(G_{\tau})_0$ its identity component.) 
Explicitly, the
matrices $M$ can be constructed as $M = \cV g_0\inv \tau_0(\cV^{-1})$;
they are gauge invariant and parameterize the coset space as $\cV$
runs through $G$. A symmetric space of the above type is called compact 
if $G$ is compact 
and non-compact otherwise; in the latter case $H$ may be both compact 
or non-compact. For non-compact symmetric spaces one can always take 
$g_0 =\1$, so that $\tau$ and $\tau_0$ coincide. 
In all cases one has $\partial_{\alpha} M M\inv = 2 D_{\alpha} \cV
\cV\inv$, with $Q_{\alpha} = \frac{1}{2}( \cV\inv \dd_{\alpha} \cV +
\tau( \cV\inv \dd_{\alpha} \cV))$. The action (\ref{sact1}) becomes
\be
S[M] = \frac{1}{8}\int d^d x \,\bra \partial^{\alpha}M M\inv, 
\,\partial_{\alpha} M M\inv \ket_{\gsf} \;,\sspace M \tau_0(M) = \pm\1\;.
\label{sact2}
\ee
Choosing local coordinates $\vp^i$ on $G/H$ the Lagrangian (\ref{sact2}) 
is just the pull-back of the line element 
\be 
\frac{1}{4} \bra d M M^{-1}, d M M^{-1} \ket_{\gsf} = \mf_{ij}(\vp) 
d\vp^i d\vp^j\,,\sspace  i,j =1, \ldots, n := \dim G/H\,,
\label{GHmetric} 
\ee
with respect to $\vp:\R^d \ra G/H$. The matrix $\mf_{ij}$ defines
a semi-riemannian metric on $G/H$ which has riemannian signature iff
$H$ is compact \cite{Loos}. By construction $\mf$ admits $\dim G$ Killing 
vectors (generating $\gsf$) $n$ of which are algebraically independent.   
Further $\mf$ enjoys the important properties 
\be
\nabla_{\!p} R_{ijkl}(\mf)= 0\,,\sspace  
R_{ij}(\mf)= \zeta_1 \,\mf_{ij}\,,
\label{R_sym}
\ee 
where the metric connection and the Riemann tensor refer to $\mf_{ij}$. 
The first property defines the larger class of locally symmetric
semi-riemannian spaces, sufficient conditions for the second
relation to hold are best formulated in terms of the Lie algebra $\gsf$.

The involution $\tau$ of $G$ induces a decomposition of the Lie algebra 
$\gsf = \msf \oplus \hsf$,
where $\hsf$ and $\msf$ are even and odd under (the differential of)
$\tau$, respectively. Further $[\hsf, \hsf] \subset \hsf,\,
[\hsf,\msf] \subset \msf,\, [\msf,\msf] \subset \hsf$. In terms of 
a basis $\{t_a\}$ of $\gsf$ with structure constants 
$[t_a, t_b] = f_{ab}^{\;\;\,c}\,t_c$ one has a corresponding decomposition
into even generators $t_{\dot{a}}, \,\dot{a} =1,\ldots, \dim \hsf$, 
and odd generators $t_{\hat{a}}, \,\hat{a} =1,\ldots, \dim \msf =n$.  
Since $\bra\cdot\,,\,\cdot\ket_{\gsf}$ is invariant under $\tau$ 
the subspaces $\hsf$ and $\msf$ are orthogonal with respect to it. 
Indices are lowered and raised with $\eta_{ab} := \bra t_a, 
t_b \ket_{\gsf}$ and its inverse. The structure constants $f_{abc}$ 
then are completely antisymmetric. The symmetric space $G/H$ is called 
indecomposable if $\msf$ does not contain a proper subspace that 
is non-singular with respect to  $\bra\cdot\,,\,\cdot\ket_{\gsf}$
and invariant under ${\rm Ad}H$. This implies that $H$ is a maximal 
subgroup of $G$. Further one then has a non-degenerate ${\rm Ad}H$ 
invariant bilinear form $\hat{\eta}_{\hat{a}\hat{b}} = 
\bra t_{\hat{a}}, t_{\hat{b}}\ket_{\gsf}$ induced on $\msf$. 
For $\exp t_{\dot{c}} \in H$ the ${\rm Ad} H$ invariance amounts to 
$\bra [t_{\hat{a}}, t_{\dot{c}}], t_{\hat{b}}\ket_{\gsf} = 
\bra t_{\hat{a}}, [t_{\dot{c}}, t_{\hat{b}}] \ket_{\gsf}$, so that,
conversely, the matrices $({\rm ad} t_{\dot{c}})_{\hat{a}}^{\;\;\hat{b}} = 
f_{\dot{c}\hat{a}}^{\;\;\;\hat{b}}$ form a subalgebra of 
${\rm so}(n,\hat{\eta})$. Note also that --  compatible with the 
complete antisymmetry of the mixed structure constants 
$f_{\dot{c}\hat{a}\hat{b}}$ -- `hatted' indices can be lowered with 
$\hat{\eta}_{\hat{a}\hat{b}}$. In this situation the Ricci tensor 
$R_{ij}(\mf)$ is non-degenerate iff $\gsf$ is semi-simple \cite{Loos}.
We consider indecomposable symmetric spaces with $\gsf$ semi-simple 
throughout. Then Eq.~(\ref{R_sym}) holds with $\zeta_1 \neq 0$; 
c.f.~below. For indecomposable symmetric spaces with riemannian signature 
a number of stronger statements hold: First the semi-simplicity of 
$\gsf$ does not have to be assumed but is a consequence. Further 
the sign $\zeta_1 >0$ then characterizes compact spaces while 
$\zeta_1 <0$ characterizes non-compact spaces \cite{Helgason}.

The curvature tensors associated with the semi-riemannian metric 
(\ref{GHmetric}) will be frequently needed. They are best computed in 
the vielbein frame.% 
\footnote{Our conventions are: 
$\nabla_i v^k = \dd_i v^k + \Gamma^k_{\;\;ij}\,v^j$, 
with $\Gamma^k_{\;\;ij} = 
\frac{1}{2}\gf^{kl}[\dd_j \gf_{il} + \dd_i \gf_{jl} - \dd_l \gf_{ij}]$.
The Riemann tensor is defined by 
$(\nabla_i \nabla_j - \nabla_j \nabla_i)v^k = 
R^k_{\;\;lij}\,v^l$, so that $R^k_{\;\;lij} = \dd_i \Gamma^k_{\;\;lj}- 
\dd_j \Gamma^k_{\;\;li} + 
\Gamma^k_{\;\;im} \Gamma^m_{\;\;\;lj} - 
\Gamma^k_{\;\;jm} \Gamma^m_{\;\;\;li}$.
The Ricci tensor is $R_{ij} = R^m_{\;\;imj}$. Let 
$e_i^{\;\,a}$ be a vielbein for $\gf_{ij} = e_i^{\;a} e_j^{\;b} \eta_{ab}$ 
with inverse $e^j_{\;a}$. The spin connection coeffcients  
$\omega_{i\;\;b}^{\;a} = e^j_{\;b} \nabla_{\!i} e_{\!j}^{\;a}$ can be read 
off from the structure equation $\dd_i e_j^{\;a} - \dd_j e_i^{\;a} = 
\omega_{i\;\;b}^{\;a} e_j^{\;b} - \omega_{j\;\;b}^{\;a} e_i^{\;b}$.  
The Riemann tensor in the vielbein frame is 
$R_{abcd} = e^i_{\;c} e^j_{\;d} [\dd_i \omega_{jba} - \dd_j \omega_{iba}  
- \omega_{i a e}\,\omega_{jb}^{\,\;\;e} + 
\omega_{ja e}\,\omega_{ib}^{\,\;\;e}].$ 
}
To introduce the vielbein we decompose the left invariant Maurer-Cartan 
form according to $\gsf = \msf \oplus \hsf$ 
\be
\cV\inv d \cV = L_i^{\;a}(\vp) t_a d\vp^i = 
L_i^{\;\hat{a}}(\vp) t_{\hat{a}} d\vp^i + 
L_i^{\;\dot{a}}(\vp) t_{\dot{a}} d\vp^i\,.
\label{MCform} 
\ee
Since $\dd_i M M\inv = 2 L_i^{\;\hat{a}}(\vp) \cV t_{\hat{a}} \cV\inv$ 
one infers from (\ref{GHmetric}) that $L_i^{\;\hat{a}}(\vp)$ is a vielbein 
for $\mf_{ij}(\vp)$, where the flat metric is $\hat{\eta}_{\hat{a}\hat{b}} 
:= \bra t_{\hat{a}}, t_{\hat{b}} \ket_{\gsf}$. In particular it follows that
there is a one-to-one correspondence between $G$-invariant semi-riemannian 
metrics on $G/H$ and ${\rm Ad} H$ invariant bilinear forms on $\msf$. 
The integrability condition for (\ref{MCform}) gives rise to 
\ba
&& \dd_i L_j^{\;\hat{a}} -\dd_j L_i^{\;\hat{a}} - 
(L_j^{\;\hat{b}} L_i^{\;\dot{c}} - L_i^{\;\hat{b}} L_j^{\;\dot{c}})
f_{\hat{b} \dot{c}}^{\;\;\,\hat{a}} =0\,,  
\nonum
&&\dd_i L_j^{\;\dot{a}} -\dd_j L_i^{\;\dot{a}} -
L_j^{\;\dot{b}} L_i^{\;\dot{c}} f_{\dot{b} \dot{c}}^{\;\;\,\dot{a}}
- L_j^{\;\hat{b}} L_i^{\;\hat{c}} f_{\hat{b} \hat{c}}^{\;\;\,\dot{a}} =0\,.
\label{MCint}
\ea 
The first equation can be interpreted as the Cartan structure equation 
for the spin connection associated with the vielbein $L_i^{\;\hat{a}}(\vp)$.
Thus $\omega_{i\;\; \hat{b}}^{\;\hat{a}} = L_i^{\; \dot{c}} 
f_{\hat{b} \dot{c}}^{\;\;\,\hat{a}}$ in the conventions declared in
the footnote. Using this and the second equation (\ref{MCint}) 
one readily computes the Riemann tensor in the vielbein frame 
\be
R_{\hat{a}\hat{b} \hat{c} \hat{d}} = f_{\dot{c} \hat{a} \hat{b}} 
\,f^{\dot{c}}_{\;\;\hat{c} \hat{d}}\,,\quad \mbox{i.e.} 
\quad t_{\hat{a}} R^{\hat{a}}_{\;\;\hat{b} \hat{c} \hat{d}} = 
-[[t_{\hat{c}},t_{\hat{d}}],t_{\hat{b}}]\,.
\label{GHRiemann}  
\ee
Converting back to the coordinate basis yields the covariant 
constancy $\nabla_{\!p} R_{ijkl}(\mf)=0$ announced in (\ref{R_sym}).
Further upon contraction one obtains from (\ref{GHRiemann}) 
$R_{\hat{a}\hat{b}} = - \frac{\varkappa}{2} \eta_{\hat{a}\hat{b}}$, if 
$\bra t_{a},\,t_b\ket_{\gsf} = \frac{1}{\varkappa} f_{ac}^{\;\;\;d} 
f_{bd}^{\;\;\;c}$. When $G$ is simple this follows because 
then $\bra\cdot\,,\,\cdot\ket_{\gsf}$ is unique up to an overall 
constant and is proportional to the Killing form; in the above 
normalization this induces $\hat{\eta}_{\hat{a}\hat{b}} = 
\frac{2}{\varkappa}f_{\dot{c}\hat{a}}^{\;\;\;\hat{c}} 
f^{\dot{c}}_{\;\;\hat{c}\hat{b}}$ and hence $\zeta_1 = - \varkappa/2$, 
in agreement with \cite{BHZJ80} in the compact case. When $G$ 
is semi-simple a choice of proportionality constant is required for each 
simple factor. The ratios of these constants can then be 
adjusted such that $R_{\hat{a}\hat{b}} \sim \eta_{\hat{a}\hat{b}}$
is recovered. Converting back to the coordinate basis results in 
$R_{ij}(\mf) = \zeta_1 \mf_{ij}$, $\zeta_1 \neq 0$, for all
indecomposable symmetric spaces with $\gsf$ semi-simple.

%%%%%%%%%%%%%%%%%%%%%%%%%%%%%%%%%%%%%%%%%%%%%%%%%%%%%%%%%%%%%%%%%%%%%%%%%
\newpage 
\newappendix{Structure of the counter terms} 

Here we derive the result (\ref{Tcounter}) on the structure of the 
$l$-loop counter terms. We repeatedly exploit two  generic properties
of them: First, they are sums of monomials built from the Riemann tensor 
of $\gf_{ij}$ and its covariant derivatives such that in each monomial 
the number of derivatives is even and at $l$-loops adds up with 
the power of the Riemann tensor according to 
\be
\mbox{power of Riemann} \;+ \;\frac{1}{2} (\# \nabla) =l\,.
\label{Tlstructure}
\ee
Second they transform as $T^{(1,l)}_{ij}(\Lambda^{-1} \gf) = 
\Lambda^{l-1} T_{ij}^{(1,l)}(\gf)$, $\Lambda \in \R$, under constant 
rescalings of the metric. For illustration and later use let us quote 
the explicit results \cite{Tseytlin87,Osb87,FokasMohamm87,Graham}
for $l \leq 3$: 
\ba
\label{T123}
T_{ij}^{(1,1)}(\gf) \is R_{ij}\;,
\nonum
T_{ij}^{(1,2)}(\gf) \is \frac{1}{4} R_{iklm}\,R_j^{\;\;klm}\;.
\\[2mm]
T_{ij}^{(1,3)}(\gf) \is \frac{1}{6} R_{imn}^{\;\;\;\;\;\;\;k} 
R_{jpqk} R^{pnmq} - \frac{1}{8} R_{iklj} R^k_{\;\;mnp}R^{lmnp} 
- \frac{1}{12} \nabla_n R_{iklm} \nabla^k R_{j}^{\;\, lmn}\,. 
\nonumber
\ea

We adopt the specifications and notations of appendix A throughout 
and begin with the 

{\bf Lemma B:} (i) Killing vectors of $\gf_{ij}$ are Killing vectors 
of $T^{(1,l)}_{ij}(\gf)$.\\  
(ii) For a symmetric space $G/H$ with $G$ simple and with metric 
$\mf_{ij}$, $i,j=1,\ldots,n$, one has 
\be
T_{ij}^{(1,l)}(\mf) = \zeta_l \,\mf_{ij} \,,\quad \zeta_l \in \R\,.
\label{T_gsym}
\ee
(iii) Let $G/H$ be a direct product of symmetric spaces as in (ii) 
with identical coefficients $\zeta_l,\,l \geq 1$. Then Eq.~(\ref{T_gsym}) 
also holds for the canonical metric on the product space.  

{\it Proof.} (i) This is an application of the well-known integrability 
conditions arising from repeated differentiation of the Killing equation
$\cL_v \gf_{ij} =0$. One obtains
\begin{equation}
\cL_v( \nabla_{i_1} \ldots \nabla_{i_d} R_{ijkl}(\gf)) =0\,,
\quad d \geq 0\,,
\label{Lie_cdRm}
\end{equation} 
for any Killing vector $v^k$ of $\gf_{ij}$ (where $R_{ijkl}(\gf)$ is 
the Riemann tensor of $\gf_{ij}$). Since the Lie derivative 
is a derivation that preserves tensor type and commutes with 
contractions, the result follows. 
%\\
(ii) For a maximally symmetric space (which is uniquely characterized
by its $n(n+1)/2$ Killing vectors and the signature) this is 
a direct consequence of (i). 
For a generic symmetric space one uses the fact that $T^{(1,l)}_{ij}(\mf)$ 
contains contractions of the Riemann tensor only, but no derivatives on
account of Eq.~(\ref{R_sym}). Further, in the vielbein frame
the Riemann tensor is constant so that $T_{ij}^{(1,l)}(\mf)$
in the vielbein frame must likewise be constant. On account of the 
$G$-invariance and the indecomposability it must be in 
one-to-one correspondence to an ${\rm Ad H}$ invariant second 
rank tensor on $\msf$ in the decomposition $\gsf = \msf \oplus \hsf$ 
of the Lie algebra. Since $G$ is simple this tensor must be proportional  
to $\hat{\eta}_{\hat{a}\hat{b}}$. Converting back to the 
coordinate basis yields the result. 
%\\
(iii) This is a direct consequence of (ii), concluding the proof of 
Lemma B.

We remark that the set of coefficients $\zeta_l,\,l \!\geq \!1$, does 
not uniquely characterize a symmetric space. A simple 
(counter-) example are maximally symmetric spaces which are uniquely 
determined by their constant sectional curvature $K$ and the 
number of positive and negative eigenvalues of the metric $\mf_{ij}$. 
From $R_{ijkl}(\mf) = K(\mf_{ik} \mf_{jl} - \mf_{il} \mf_{jk})$ 
one sees that the curvature scalars $\zeta_l$ depend on $K$ and the 
dimension $n$ but on the signature of the metric at most through 
the sign of $K$. For example one computes from (\ref{T123}) 
\begin{equation}
\zeta_1 = K(n-1) \;,\quad 
\zeta_2 = \frac{K^2}{2} (n-1) \;,\quad 
\zeta_3 = \frac{K^3}{12}\big( (n+1)^2 -4 \big)\;,
\label{zeta_gmax}
\end{equation}
in agreement with \cite{BrezHik78,Hikami81}.

The target space metric (\ref{Tmetric}) of course is not that of a 
symmetric space. In order to be able to apply the Lemma we need
\begin{subeqnarray}
&& T^{(1,l)}_{ij}(\gf) = \frac{1}{h^{l-1}} \, T^{(1,l)}_{ij}(\mf)\,,
\sspace \;\;\,  i,j = 1,\ldots, n\,,\\
&& T^{(1,l)}_{\sigma\rho}(\gf) =0\,,\quad 
T^{(1,l)}_{\sigma i}(\gf) =0\,,\quad 
T^{(1,l)}_{\rho i}(\gf) =0\,.
\label{Tblock}
\end{subeqnarray}
To show this one consecutively computes the Christoffel symbols, the Riemann 
tensor and its covariant derivative for the metric (\ref{Tmetric}). 
The independent non-vanishing components are:  
\begin{subeqnarray} 
&& \Gamma^k_{\;ij} = \Gamma^k_{\;ij}(\mf)\,,
\quad 
\Gamma^k_{\;i\rho} = \frac{1}{2} \dd_{\rho}\! \ln h \,\delta_i^k\,,
\quad 
\Gamma^{\rho}_{\;\rho\rho} = \dd_{\rho} \ln (h/\rho)\,.
\nonumber\\
&& \Gamma^{\sigma}_{ij} = - \frac{1}{2b} \rho \dd_{\rho} \ln h \, 
\mf_{ij}\,,\quad \Gamma^{\sigma}_{\;\rho\rho} = - \frac{a}{2b\rho}
\dd_{\rho} \ln h\,,
\\
&& R_{ijkl} = h(\rho)\, R_{ijkl}(\mf) \,,\quad R_{\rho  i \rho j } =
\frac{h(\rho)}{\rho^2} S_1(\rho)\; \mf_{ij} \,,
\\
&& \nabla_{\!\rho} R_{ijkl} = 2 \nabla_{\!i} R_{\rho jkl} =  
- \dd_{\rho} h \,R_{ijkl}(\mf) \,,
\quad \nabla_{\!\rho} R_{\rho i \rho j} = \frac{h^3}{\rho^2} \dd_{\rho} 
\Big(\frac{S_1}{h^2} \Big) \, \mf_{ij} \,,
\label{Rmg} 
\end{subeqnarray}      
with $S_1(\rho)$ as in (\ref{Sl}) and $i,j,k,l \in \{1,\ldots,n\}$. 
All components not related to these by symmetries of the Riemann tensor 
vanish. In Eq.~(\ref{Rmg}c) the covariant constancy (\ref{R_sym}) 
enters. 

For $l \leq 3$ one can now verify the asserted structure (\ref{Tcounter})
by explicit computation, using Eqs.~(\ref{R_sym}), (\ref{T123}) and  
Lemma B. In particular one thereby obtains the expressions for 
$S_l(\rho),\,l=1,2,3$, anticipated in (\ref{Sl}).

Proceeding with the general analysis one shows by induction from 
(\ref{Rmg}) and (\ref{R_sym})
\be
\nabla_{i_1} \ldots \nabla_{i_d} R_{ijkl} \neq 0 \quad 
\mbox{only if} \quad 
\left\{
\begin{array}{l} \# \rho \;\mbox{indices}\; \neq 0,\\ 
\# \rho \;\mbox{indices} \;+ \;\# \;
\mbox{derivatives} = \mbox{even} \,.
\end{array}
\right.
\label{Rm_derivatives} 
\ee
In particular all components containing one or more `lower' $\sigma$ index
vanish. Further $\rho$ cannot appear as a summation index since an  
`upper' $\rho$ index amounts to a `lower' $\sigma$ index. 
Combining this with (\ref{Rmg}b) establishes Eq.~(\ref{Tblock}a)      
and $T^{(1,l)}_{\sigma \rho}(\gf) = T^{(1,l)}_{\sigma\,i}(\gf)=0$. 

To show that also  $T^{(1,l)}_{\rho\,i}(\gf)$ vanishes for $i =1,\ldots,n$, 
one can proceed as follows. Since $\rho$ cannot appear as a summation index 
the monomials in $T^{(1,l)}_{\rho\,i}(\gf)$ must have the following form
\ba
&& \nabla_{i_1} \ldots \nabla_{i_d} R_{i_{d+1} \ldots i_{d+4}} 
\;Q(R)_i^{i_1.. \hat{i}_r.. i_{d+4}}\;,
\quad \quad \mbox{with} \;\; \rho = i_r, \quad \mbox{or}\,,
\nonum
&& \nabla_{i_1} \ldots \nabla_{i_d} R_{i_{d+1} \ldots i_{d+4}} 
\;Q(R)^{i_1.. \hat{i}_r..\hat{i}_s .. i_{d+4}}\;,
\quad \mbox{with} \;\;\rho = i_r, i = i_s\,,
\label{Trhoi}
\ea 
for some $r,s \in\{1,\dots, d\!+\!4\},\,d \geq 0$, and the summation is 
over all but the `hatted' indices, which are omitted. Here $Q(R)$ is 
a tensor which by 
(\ref{Rm_derivatives}) cannot contain derivatives of the Riemann 
tensor. Thus all derivatives must be carried by the first term and 
$d$ in (\ref{Trhoi}) equals the total number of derivatives. On account of 
(\ref{Tlstructure}) $d$ must be even, and so must be the number of 
$\rho$ indices in view of Eq.~(\ref{Rm_derivatives}). However this 
contradicts Eq.~(\ref{Trhoi}) which allows only a single $\rho$ index. 
Hence no non-zero candidate monomial exists and we infer
$T^{(1,l)}_{\rho i}(\gf) =0$. This concludes the derivation of 
Eq.~(\ref{Tblock}). We note that the result hinges on the 
covariant constancy in Eq.~(\ref{R_sym}); in particular the block 
diagonal form of the counter terms is not a trivial consequence of 
the block diagonal form of the metric (\ref{Tmetric}).

The rest is straightforward: Applying part (ii) of Lemma B to 
Eq.~(\ref{Tblock}a) gives $T^{(1,l)}_{ij}(\gf) = h^{1-l} 
\zeta_l\, \mf_{ij}$. The remaining matrix element $T^{(1,l)}_{\rho\rho}(\gf)$
has to scale with weight $1\!-\!l$ under constant rescalings of $h$. 
Extracting an explicit power of $h$ one can write 
$T^{(1,l)}_{\rho\rho}(\gf) = n h^{1-l} S_l(\rho)/\rho^2$, for some  
differential polynomial $S_l$ in $h$ invariant under 
constant rescalings of $h$. For constant $h$ the metric (\ref{Tmetric}) 
is that of a direct product of the (irreducible) symmetric space $G/H$ 
with $\R^{1,1}$, and it is not hard to see that $S_l(\rho)$ then vanishes. 
This concludes the derivation of Eq.~(\ref{Tcounter}).

\newpage  
%%%%%%%%%%%%%%%%%%%%%%%%%%%%%%%%%%%%%%%%%%%%%%%%%%%%%%%%%%%%%%%%%%%%%%%%%

\newappendix{Proof of UV stability}

Here we establish the result on the UV stability described in 
section 4. It is sufficient to consider the case 
$\zeta_1 <0$ and $t \geq 0$. The flow equation (\ref{linflow}) is
invariant under $t \ra -t$, $\zeta_l \ra -\zeta_l,\,l \geq 1$. 
Thus if indeed, as the theorem asserts, $\zeta_1 <0$ implies decay 
of the perturbations for $t \ra \infty$ irrespective of the values 
of $\zeta_2,\zeta_3,\ldots$, a positive $\zeta_1$ will imply decay 
for $t\ra -\infty$, i.e.~(formal) infrared stability of the fixed-point. 
We therefore assume $\zeta_1<0$ throughout this appendix and set 
$t_1 := -\zeta_1 t \geq 0$. In the following we first outline the 
derivation of Eq.~(\ref{slrecursion})  and then show that the solutions 
recursively constructed thereby have the announced properties.

Since integro-differential equations are cumbersome we 
convert Eqs.~(\ref{slflow}) into a system of partial differential 
equations 
\be
\dd_t \ssbar_l + \zeta_1 \dd_{\varrho} \ssbar_l =
\varrho \dd_{\varrho} [R_l/\varrho]\,,
\quad l \geq 1\,,
\label{tlflow} 
\ee 
where $\ssbar_l = \varrho \dd_{\varrho} (\sbar_l/\varrho)$. The general 
solution of Eq.~(\ref{tlflow}) at given $l$ is a sum of the generic 
solution to the homogeneous equation and a particular solution of the 
inhomogeneous equation. The former can be described in terms of a (smooth) 
function $r_l$ of one variable as $-r_l(\varrho - \zeta_1 t)$.  
Regarding $\varrho\dd_{\varrho} [R_l/\varrho]$ as an (in principle) 
known function of $\varrho$ and $t$ the full solution of (\ref{tlflow}) 
is  
\be
\ssbar_l(\varrho,t) =
- \, r_l(\varrho -\zeta_1 t) + \int_0^t ds 
\Big(\varrho \dd_{\varrho} [\varrho^{-1} R_l]\Big)
(\varrho - \zeta_1 s,\,t-s) \,.
\label{Slsol} 
\ee
The solutions for $\sbar_l$ are obtained by integration
\be
\sbar_l(\varrho,t) = - \varrho \int_{\varrho}^{\infty} 
\frac{du}{u} \ssbar_l(u,t)\,, 
\label{slRlsol}
\ee
where the upper integration boundary is chosen with regard to 
the boundary condition at $\varrho = \infty$ aimed at (though of 
course it does not in itself entail it). Inserting (\ref{Slsol}) 
into (\ref{slRlsol}) with 
\be
\varrho\dd_{\varrho} [R_l/\varrho] = \sum_{k=2}^l \varrho^{-k-1}
[ \beta_k \varrho^2  \dd_{\varrho}^2 +(\alpha_k - k\beta_k)
\varrho \dd_{\varrho} - (\gamma_k\!+\!(k\!+\!1)\alpha_k) ]
\sbar_{l+1-k}(\varrho,t)\,,
\label{Rlder} 
\ee
and performing some integrations by parts, one 
arrives at the recursive solution Eq.~(\ref{slrecursion}).

A simple consequence is that there can be at most one solution of 
(\ref{slflow}) with given (smooth) initial data $\sbar_l(\varrho,t\!=\!0)$.
From Eq.~(\ref{slrecursion}) one infers:   
$\sbar_l(\varrho,t) = 0$ for all $t$ if $r_k(\varrho) = 
- \varrho \dd_{\varrho} [ \sbar_k(\varrho,t\!=\!0)/\varrho] =0$ for
$k \leq l$ $(*)$. Thus if there were two distinct solutions with the same 
initial data, their difference $\sbar^{\rm diff}_l(\rho,t)$ would solve 
(\ref{slflow}) by linearity. Then $(*)$ applies and entails
$\sbar^{\rm diff}_l(\rho,t) \equiv 0$, i.e.~both solutions coincide.    
A similar argument shows that the system (\ref{slflow}) does 
not admit `static' solutions satisfying the required boundary 
conditions. Indeed, a $t$-independent solution 
for $\sbar_1$ is proportional to $\varrho \ln \varrho$ which 
satisfies the boundary condition only if the proportionality 
constant vanishes. Hence $\sbar_1 =0$ and $R_2[\sbar_1] =0$,
so that the $l=2$ equation now gives $\sbar_2 =0$. Since  
$R_l$ vanishes if $\sbar_1, \ldots, \sbar_{l-1}$ vanish
the absence of static solutions follows by iteration. Finally it is 
also clear that the solutions produced by (\ref{slrecursion}) from 
smooth initial data will the smooth in both $t$ and $\varrho$ 
(for $\varrho$ bounded away from zero) as long as the integrals 
involved converge absolutely. The latter will be a byproduct of the 
inductive bounds shown below.

For any smooth function $r(u)$ on $\R^+$ satisfying $u\,r(u) \ra 0$ 
for $u \ra \infty$ we set
\be 
\sbar^{\rm hom}(\varrho, t) = \varrho \int_{\varrho}^{\infty} 
\frac{du}{u} r( u - \zeta_1 t)\,, 
\label{slhom}
\ee  
which describes the homogeneous parts of the solutions (\ref{slrecursion}).
They are readily seen to satisfy both the premise and the conclusion 
of the theorem: By definition $M_x := \max_{u \geq x > 0} u |r(u)|$ 
is a finite positive constant, decreasing with increasing x such that 
$M_{\infty}=0$. For $t_1 = -\zeta_1 t >0$ one verifies the bounds 
\be
|\sbar^{\rm hom}(\varrho,t)| \leq M_{\varrho + t_1} \;,
\sspace
|\varrho \dd_{\varrho} \sbar^{\rm hom}(\varrho,t)| 
\leq 2 M_{\varrho + t_1} \;,
\label{slhom_bounds}
\ee
using $ \varrho \dd_{\varrho} \sbar^{\rm hom} = \sbar^{\rm hom} - 
\varrho r(\varrho + t_1)$ in the second case. 
For all $t_1 \geq 0$ then $M_{\varrho + t_1} \leq M_{\varrho}$,
so that for $\varrho$ large the bound is uniform in $t_1$.
Likewise for $t_1$ large one has $M_{\varrho + t_1} \leq M_{t_1}$ 
for all $\varrho >0$, giving a bound uniform in $\varrho$.     
\bigskip

{\bf Strategy for inductive bounds:} The aim in the following is 
to show by induction on $l$ that the solutions recursively produced by 
(\ref{slrecursion}) indeed have the announced uniform decay properties, 
both for $\varrho \ra \infty$ (boundary condition) and for 
$t \ra \infty$ (renormalization group dynamics). We decompose each 
$\sbar_l$ as 
$\sbar_l = \sbar_l^{\rm hom} + \sbar_l^{\rm inh}$, where 
$\sbar_l^{\rm hom}$ is of the form (\ref{slhom}) with $r=r_l$ 
and $\sbar^{\rm inh}_l$ is the remainder in (\ref{slrecursion}).
The homogeneous parts are already known to have the desired 
properties. For the inhomogeneous parts we seek to establish uniform 
bounds for each of the terms on the right hand side of 
\be
|\sbar_l^{\rm inh}(\varrho,t)| \leq 
\sum_{k=2}^l \int_0^t ds \Big| F_k[\sbar^{\rm hom}_{l+1 -k}] \Big| + 
\sum_{k=2}^l \int_0^t ds \Big| F_k[\sbar^{\rm inh}_{l+1 -k}] \Big|\,.
\label{strat1}
\ee 
It turns out that for all but the $|F_2[\sbar^{\rm inh}_{l-1}]|$ 
term this is relatively straightforward. When estimated using only 
the induction hypothesis for $\sbar^{\rm inh}_{l-1}$ (i.e.~that
$\sbar^{\rm inh}_{l-1}$ decays for $t\ra \infty$ uniformly in $\varrho$,
but the decay may be arbitrarily soft) this term
seems to diverge logarithmically for $t \ra \infty$ at fixed 
$\varrho$. In order to overcome this problem we establish in Lemma 2 
below that the inhomogeneous parts $\sbar^{\rm inh}_{l+1-k}$ actually 
decay faster, like an inverse power of $t$. For 
the homogeneous parts, in contrast, no such result holds and they may 
decay arbitrarily soft for $t \ra \infty$. Clearly both statements are 
compatible only if the linear functional of the $\sbar^{\rm hom}_{l+1 -k}$  
appearing in (\ref{strat1}) decays like an inverse power of $t$. 
This is what we show first:

{\bf Lemma C1:} Let $\sbar^{\hom}(\varrho,t)$ be of the 
form (\ref{slhom}) for some smooth function $r(u)$ satisfying $u\,r(u) \ra 
0$, for $u \ra \infty$. Then there exists $0< \eps <1$ 
such that for all $k =2,\ldots, l$
\be
t^{\eps} \int_0^t ds F_k[\sbar^{\rm hom}](\varrho,t;s) 
\rra 0 \quad \mbox{for} \quad t \ra \infty\,,
\label{Lemma1}
\ee
where the convergence is uniform in $\varrho$, for all $\varrho$ bounded 
away from zero. For $\varrho \ra \infty$ the left hand side also vanishes 
uniformly in $t$.

{\it Proof.} For $k \geq 2$ we rewrite the integrand as follows 
\ba 
F_k[\sbar^{\rm hom}] \is \varrho \int_{\varrho}^{\infty} du 
\frac{P_k(u,s)}{u - \zeta_1 s}  r(u + t_1) 
- \frac{\beta_k}{(\varrho - \zeta_1 s)^{k-1}} r(\varrho +t_1)  
\nonum
& + &  
\frac{\sbar^{\rm hom}(\varrho - \zeta_1 s, t-s)}{\varrho - \zeta_1 s} 
\left[ \frac{\alpha_k + \beta_k - \beta_k \zeta_1 s/\varrho}% 
{(\varrho - \zeta_1 s)^{k-1}} - \varrho P_k(\varrho,s)\right]\,,   
\nonum
P_k(\varrho,s) \is - \int_{\varrho}^{\infty} \!du\, 
\frac{a u + s b + c s^2/u}{u^2 (u - \zeta_1 s)^k}\,.  
\label{Fkhom1}
\ea
The constants $a,b,c$ entering $P_k$ can be read off from 
(\ref{slrecursion}c) but except for $k=2$ their values are inessential. 
For $k\geq 3$ we shall later use the function symbol $P_k$ also 
when the values are different. 

For $k=2$ specifically the relations $\alpha_2 = \beta_2 = -\gamma_2/4$ 
among the coefficients in (\ref{Rk1coeffs}) imply that
(\ref{Fkhom1}) simplifies as follows
\ba
\label{F2hom1}   
&& \int_0^t ds F_2[\sbar^{\rm hom}](\varrho,t;s) 
\\[1mm]
&& \quad = - \frac{2 \zeta_2}{\zeta_1} 
\varrho \int_{\varrho}^{\infty} \!du \,r(u + t_1) \dd_u 
\Big[ \frac{1}{u} \ln\Big(1 + \frac{t_1}{u} \Big) \Big] 
- \frac{2 \zeta_2}{\zeta_1} r(\varrho + t_1) \ln(1 + t_1/\varrho) \,,
\nonumber
\ea
using 
\be
P_2(\varrho,s) = \frac{2\zeta_2}{\varrho^2} 
\frac{\zeta_1 s - 2 \varrho}{\varrho - \zeta_1 s}\,,\sspace 
\int_0^t ds P_2(\varrho,s) = -\frac{2\zeta_2}{\zeta_1} 
\dd_{\varrho}\Big[\frac{1}{\varrho} 
\ln \Big( 1 + \frac{t_1}{\varrho} \Big)\Big]\,. 
\label{P2}  
\ee
A straightforward estimate then is 
\be
\int_0^t ds \Big| F_2[\sbar^{\rm hom}](\varrho,t;s) \Big| 
\leq \frac{M_{\varrho + t_1}}{\varrho}    
\Big|\frac{4 \zeta_2}{\zeta_1} \Big|\,
\frac{\ln(1 + t_1/\varrho)}{1+ t_1/\varrho}\,. 
\label{F2hom3}
\ee
Here $M_x := \max_{u \geq x > 0} u |r(u)|$ as before.
Observing that for $0< \eps <1$ and $x \geq 0$ one has 
$(1+x)^{\eps -1} \ln(1 + x)\leq [(1-\eps)e]^{-1}$, the required uniform 
bounds follow: For all $t_1 \geq 0$ and $\varrho$ large have 
{\it lhs} $\leq |4 \zeta_2/(e\zeta_1)|\,M_{\varrho}/\varrho$. 
Likewise for given $\varrho_0 > 0$ and $t_1$ large have 
$t_1^{\eps} \times$ {\it lhs} $\leq |4 \zeta_2/((1-\eps)e\zeta_1)|\,
M_{t_1}/\varrho_0$, for all $\varrho > \varrho_0$.

For $k \geq 3$ we prepare the bounds%
\footnote{We use $C,C_1,C_2,\ldots$, as placeholders for positive 
$k$-dependent constants; in different formulas usually different 
constants appear.} 
\ba 
\label{Fkhom2}
&& \Big| \frac{\sbar^{\rm hom}(\varrho - \zeta_1 s, t-s)}%
{\varrho - \zeta_1 s}\Big| \leq \frac{M_{\varrho + t_1}}{t_1} 
\ln\Big(1 + \frac{t_1}{\varrho}\Big)\,, 
\nonumber \\[3mm]
&& \int_0^t ds \,\Big| \frac{P_k(\varrho,s)}{\varrho - \zeta_1 s} \Big| 
\leq \frac{C}{\varrho^k}\left[1 - \Big(\frac{\varrho}{\varrho + t_1}\Big)^{k-1}
\right]\,,\sspace \quad \;k\geq 3\,,
\\[3mm] 
&& \int_0^t ds \,|P_k(\varrho,s)| \leq
\left\{ 
\begin{array}{ll} 
{\displaystyle \frac{C_1}{\varrho^2}\left[
1+ C_2\ln\Big( 1 + \frac{t_1}{\varrho} \Big) \right]\,,} 
\quad & k=3\,,\\[4mm]  
{\displaystyle \frac{C}{\varrho^{k-1}} }\,,\quad & k\geq 4\,. 
\end{array} 
\right.
\nonumber
\ea 
A simple computation then gives 
\ba
\label{Fkhom3}
&& \int_0^t ds \Big| F_k[\sbar^{\rm hom}](\varrho,t;s) \Big| 
 \leq  \frac{M_{\varrho + t_1}}{\varrho + t_1}
\frac{C_1}{\varrho^{k-2}} 
\\[2mm]
&&\sspace  + \frac{M_{\varrho + t_1}}{t_1} 
\frac{C_2}{\varrho^{k-2}} \ln\Big(1 + \frac{t_1}{\varrho}\Big) 
\times \left\{ 
\begin{array}{ll} 
{\displaystyle 1 + C_3 \ln\Big( 1 + \frac{t_1}{\varrho} \Big) } 
\quad & k=3\,,\\[4mm]  
1 \quad & k\geq 4\,. 
\end{array} 
\right.
\nonumber
\ea 
From here the required uniform bounds follow. This concludes the 
proof of Lemma C1. 
 
Having the homogeneous part in (\ref{strat1}) under control one can bound 
the inhomogeneous part with an inductive argument.

{\bf Lemma C2:} For $l \geq 2$ let $\sbar^{\rm inh}_l$ be the 
inhomogeneous part of the solution recursively generated by 
Eq.~(\ref{slrecursion}). Then there exists $0< \eps <1$ such that
\be
t^{\eps}\, \Big| \sbar^{\rm inh}_l(\varrho,t) \Big| \rra 0
\quad \mbox{for} \quad t \ra \infty\,,
\label{Lemma2}
\ee
uniformly in $\varrho$, for $\varrho$ bounded away from zero. 

{\it Proof.} For $l=2$ the result follows
from Eq.~(\ref{F2hom3}) with $\sbar^{\rm hom} = \sbar_1$. For $l \geq 3$ 
we use Eq.~(\ref{strat1}) where the homogeneous part is taken care 
of by Lemma 1 and proceed by induction. Assuming that $\sbar^{\rm inh}_2,
\ldots,\sbar^{\rm inh}_{l-1}$ are already known to satisfy 
Eq.~(\ref{Lemma2}) we have to bound $\int_0^t ds | 
F_k[\sbar^{\rm inh}_{l+1 -k}]|$ in Eq.~(\ref{strat1}). 
For $k =2,\ldots, l-1$, define 
\be
M_k(\varrho,t) = \max_{\begin{array}{c} 
\scriptstyle{0 \leq s \leq t} \\[-3mm] \scriptstyle{\rho \leq u < \infty} 
\end{array}} \!\!\!
\Big\{ |\sbar^{\rm inh}_k(u - \zeta_1 s, t-s)|,\; 
|u \dd_u \sbar^{\rm inh}_k(u - \zeta_1 s, t-s)| \Big\}\,. 
\label{Maxus} 
\vspace{-3mm} 
\ee 
By the induction hypothesis $M_k(\varrho,t)$ is finite for all 
$\varrho,t>0$, vanishes for $\varrho \ra \infty$ 
uniformly in $t$, and satisfies $t^{\eps} M_k(\varrho,t) \ra 0$, 
as $t \ra \infty$ uniformly in $\varrho$ for all $\varrho$ bounded 
away from zero. This stronger induction hypothesis now makes the 
rest of the argument straightforward. 

For $k =2, \ldots ,l-1$, one verfies 
\be
|F_k[\sbar^{\rm inh}_{l+1-k}]| \leq M_{l+1-k}(\varrho,t) \left\{ 
\varrho |P_{k+1}(\varrho,s)| 
 + \frac{|\alpha_k| + |\beta_k|}{\varrho(\varrho -\zeta_1 s)^{k-1}}
\right\}\,.
\label{Fkinh1}
\ee
Using the last Eq.~in (\ref{Fkhom2}) this yields 
\be
\int_0^t ds \Big|F_k[\sbar^{\rm inh}_{l+1-k}](\varrho,t;s)\Big| \leq 
M_{l+1-k}(\varrho,t) \frac{C_1}{\varrho^{k-1}} 
\left\{ \begin{array}{ll} 
{\displaystyle 1 + C_2 \ln\Big( 1 + \frac{t_1}{\varrho}\Big)} 
\,, \quad & k =2\,,
\\
1\,, & k \geq 3\,.
\end{array}
\right.
\label{Fkinh2}  
\ee
From here the required uniform bounds readily follow, completing the 
proof of Lemma C2. Lemmas C1,C2 and Eq.~(\ref{slhom_bounds}) imply the 
theorem.

%%%%%%%%%%%%%%%%%%%%%%%%%%%%%%%%%%%%%%%%%%%%%%%%%%%%%%%%%%%%%%%%%%%%%%%%%
\newpage 
%\bibliographystyle{else}\bibliography{bib}

%%%%%%%%%%%%%%%%%%%%%%%%%%%%%%%%%%%%%%%%%%%%%%%%%%%%%%%%%%%%%%%%%%%%%%%%%%%%%

\end{document}